\documentclass[
reprint,
superscriptaddress,
amsmath,amssymb,
prb,
]{revtex4-2} 

\usepackage{graphicx}
\usepackage{dcolumn}
\usepackage{bm}
\usepackage{xcolor}
\usepackage{float}
\usepackage{textcomp}
\usepackage{siunitx}
\usepackage{booktabs}
\usepackage[version=4]{mhchem}

\begin{document}  
	
\title{Interaction-Driven Chern Insulator at Zero Electric Field \\in ABCB-Stacked Tetralayer Graphene}%
    
\author{Yulu Ren}
\affiliation{Key Laboratory of Artificial Structures and Quantum Control (Ministry of Education), School of Physics and Astronomy, Shanghai Jiao Tong University, Shanghai 200240, China}
\author{Yang Shen}
\affiliation{School of Physical Science and Technology, ShanghaiTech University, Shanghai 201210, China}
\author{Chengyang Xu}
\affiliation{Key Laboratory of Artificial  Structures and Quantum Control (Ministry of Education), School of Physics and Astronomy, Shanghai Jiao Tong University, Shanghai 200240, China}
\author{Wanfei Shan}
\affiliation{College of Letters and Science, University of California, Los Angeles, California 90095, USA}
\author{Weidong Luo}
\email{wdluo@sjtu.edu.cn}
\affiliation{Key Laboratory of Artificial Structures and Quantum Control (Ministry of Education), School of Physics and Astronomy, Shanghai Jiao Tong University, Shanghai 200240, China}

\begin{abstract}
ABCB-stacked tetralayer graphene, with intrinsic spontaneous polarization, offers a unique platform to explore electron correlation effects, whose interplay with spin-orbit coupling may engender topological phases.
Here, employing a $\mathbf{k}\cdot\mathbf{p}$ model with self-consistent Hartree-Fock calculations, we investigate its electronic ground states.
Remarkably, we find that the intrinsic polarization, in conjunction with strong interactions ($U=8 \text{ eV}$) and SOC, is sufficient to drive a $C=3$ quantum anomalous Hall state, obviating the need for an external electric field typical in ABCA stacks.
Conversely, at moderate interactions ($U=6 \text{ eV}$), a minimal electric field is necessary.
Furthermore, calculations predict other correlation-driven metallic phases such as quarter- and three-quarter-filled states.
These results establish that the synergy of intrinsic polarization, correlations, and SOC governs the rich topological phenomena, suggesting ABCB-stacked graphene as a highly tunable platform for exploring emergent topological phenomena.
\end{abstract}
\maketitle

\section{\label{sec:level1}INTRODUCTION}
Electron correlations give rise to a diverse array of unconventional quantum states extending beyond the single-particle picture, consequently garnering substantial scientific interest\cite{bansil_colloquium_2016,cao_correlated_2018,lu_superconductors_2019,tran_evidence_2019,cao_unconventional_2018}.
Such interactions manifest diversely, yielding canonical correlated states such as Mott insulators\cite{chen_evidence_2019,ma_metallic_2016,regan_mott_2020,po_origin_2018}, unconventional superconductivity\cite{sato_topological_2017,ortiz_cs_2020,park_tunable_2021,po_origin_2018,balents_superconductivity_2020}, and charge density waves\cite{ugeda_characterization_2016,li_controlling_2016,chen_double_2021,ma_metallic_2016}. Furthermore, the pivotal role of correlation in engendering novel topological states, notably Chern insulators, is increasingly evident in recent years\cite{chen_tunable_2020,zhang_nearly_2019,polshyn_electrical_2020,wang_fractional_2024}.
Graphene and other 2D van der Waals materials, when prepared with high quality, atomic precision, and tunability, are exceptional for studying electron correlations and resultant topological states\cite{choi_synthesis_2010,park_growth_2010,gupta_recent_2015,akinwande_review_2017,bhimanapati_recent_2015}.
For instance, engineered flat bands originating from moiré superlattices have been shown to markedly enhance electron interactions, leading to the phenomena such as unconventional superconductivity and Mott insulating phases\cite{wu_chern_2021,das_symmetry-broken_2021,liu_quantum_2019}. 
Moreover, interaction-driven spontaneous symmetry breaking in these systems can give rise to a rich spectrum of other novel quantum phases, including integer and fractional quantum anomalous Hall effects, tunable valley- and spin-polarized states, and orbital ferromagnetism\cite{lu_superconductors_2019,liu_quantum_2019,han_orbital_2023,chen_tunable_2020,lu_fractional_2024}.
    
Rhombohedral-stacked multilayer graphene (nRG), with its $E \sim k^N$ dispersion and high density of states (DOS) near charge neutrality point (CNP), facilitates intrinsic interaction phenomena\cite{koshino_multilayer_2013,castro_neto_electronic_2009,geim_graphene_2009}.
For instance, in 3RG, an external perpendicular electric fields ($E$) induce interaction-driven quarter-metal states, and spin-orbit coupling (SOC) can promote superconductivity\cite{yang_impact_2025,patterson_superconductivity_2025,zhang_enhanced_2023}.
In 4RG and 5RG, interactions spontaneously yield a layer antiferromagnetic (LAF) ground state\cite{liu_spontaneous_2024}. This state, in turn, serves as a platform for realizing integer Chern insulators by introducing SOC with proximitized $\text{WSe}_2$ and applying perpendicular electric field and magnetic fields ($B$)\cite{li_even-denominator_2017,sha_observation_2024,makov_flat_2024,zhou_fractional_2024,han_correlated_2024,lu_fractional_2024,winterer_ferroelectric_2024,han_orbital_2023}.
This highlights that interaction-mediated spontaneous symmetry breaking is crucial for Chern insulator phases.
Therefore, a pivotal question remains: can intrinsic spontaneous polarization, together with correlations, function as a substitute for an electric field, either partially or entirely, to drive the formation of Chern insulators with $\text{WSe}_2$-induced SOC?  

ABCB-stacked tetralayer graphene (ABCB) represents a compelling candidate material.
Notably, it is the sole system reported to exhibit spontaneous polarization among configurations with four or fewer layers\cite{wirth_experimental_2022}.
From the perspective of stacking stability, recent DFT calculations\cite{paul_nery_ab-initio_2021} show that, with the stable ABAB taken as the reference zero, the ground-state energy of the metastable ABCA is only 0.007 meV/carbon atom higher, while that of the ABCB is just 0.015 meV/carbon atom higher. Consistently, another study\cite{yang_atypical_2023} reports that the energy differences among different four-layer stacking configurations are within 0.038 meV/carbon atom. Furthermore, it predicts a low energy barrier of less than 5 meV/unit cell (0.625 meV/carbon atom) for the interlayer sliding transition between ABCB and BCBA configurations\cite{yang_atypical_2023}. These results indicate that the ABCB in four-layer graphene is metastable, with its energy differing only slightly from that of the stable configuration, and that the sliding barriers between alternative stackings are relatively low.

Furthermore, the DOS peak in ABCB near the CNP is found to be comparable to that in its ABCA-stacked counterpart (ABCA) according to previous theoretical studies~\cite{fischer_spin_2024} and our own calculations~\cite{unpublished}.
Indeed, the same study highlights the potential for superconductivity induced by spin and charge fluctuations upon the inclusion of both long- and short-range interactions\cite{fischer_spin_2024}.
However, current research on ABCB predominantly concentrates on its distinctive ferroelectric properties\cite{yang_atypical_2023}.
Experimentally, hysteresis loops have been observed in hexagonal boron nitride (hBN)-encapsulated ABCB devices during fast sweeps of both top and back gate voltages.
This observation supports that the ferroelectricity is an intrinsic characteristic of the material, rather than an artifact of moiré potentials\cite{singh_stacking-induced_2025}.
Collectively, while existing studies have begun to explore its polarization phenomena and potential electron correlation effects, the intricate interplay between the intrinsic spontaneous polarization and electronic correlations in ABCB remains a compelling open question. 
To elucidate the subtle interplay between intrinsic polarization and electronic correlations in ABCB, we employ a refined $\mathbf{k}\cdot\mathbf{p}$ effective model with self-consistent Hartree-Fock calculations. 
Our results demonstrate that strong electron interactions ($U = 8 \, \text{eV}$), in conjunction with SOC, drive the system into a quantum anomalous Hall (QAH) state characterized by a Chern number $C=3$, in the absence of an electric field.
Conversely, with moderately reduced interactions ($U = 6 \, \text{eV}$), the system requires a small external electric field to transition into the Chern insulator state.
Indeed, the interplay between the interactions $U$ and the spontaneous polarization co-determines the Chern topological phase and its dependence on electric field $E$.
Furthermore, our calculations reveal the possibility of other correlation- and SOC-induced symmetry-broken metallic states, including quarter- and three-quarter-filled phases.

\section{\label{sec:level1}RESULTS} 

\subsection{\label{subsec:level2}Model Hamiltonian} 
To investigate the topological phases in ABCB-stacked graphene driven by electron correlations, we first establish a theoretical framework that accurately captures its low-energy physics. In this section, our central goal is to construct a comprehensive effective Hamiltonian. We will detail each term of this Hamiltonian, the physics it describes, and the selection of relevant parameters. The total Hamiltonian, $\mathcal{H}_{\text{tot}}$, encompasses the effective non-interacting Hamiltonian ($\mathcal{H}_{\text{eff}}$), spin-orbit coupling ($\mathcal{H}_{\text{soc}}$), and electron-electron interactions ($\mathcal{H}_{\text{int}}$), and is expressed as:
\begin{equation}
	\mathcal{H}_{\text{tot}} = \mathcal{H}_{\text{eff}} + \mathcal{H}_{\text{soc}} + \mathcal{H}_{\text{int}}.
	\label{eq:H_total_revised}
\end{equation}

The effective non-interacting Hamiltonian, $\mathcal{H}_{\text{eff}}$, is given in the second quantization formalism by:

\begin{equation}
	\mathcal{H}_{\text{eff}} = \sum_{\mathbf{k}, \tau, s} f_{\mathbf{k}, \tau, s}^\dagger h_{\text{eff},\tau,s}(\mathbf{k}) f_{\mathbf{k}, \tau, s},
	\label{eq:Heff_phi}
\end{equation}
where $f_{\mathbf{k},\tau,s} = (f_{A_1,\mathbf{k},\tau,s}, f_{B_1,\mathbf{k},\tau,s}, \dots, f_{B_N,\mathbf{k},\tau,s})^T$ is the annihilation operator for an electron with momentum $\mathbf{k}$, valley index $\tau$, and spin $s$, with components corresponding to the different sublattices (see Fig. \ref{fig:a3} for site index).

The effective Hamiltonian matrix $h_{\text{eff},\tau,s}(\mathbf{k})$ includes contributions from both intra- and inter-layer hopping ($h_{\text{t},\tau}$) and an on-site potential ($V_{\text{os}}$).
\begin{equation}
h_{\text{eff},\tau,s}(\mathbf{k}) = h_{\text{t},\tau}(\mathbf{k}) + V_{\text{os}},
\end{equation}
The term $V_{os}$ introduces on-site potentials for each site within the unit cell, which are determined from DFT calculations (see Appendix \ref{APP_A} for details) and given by:
$$V_{\text{os}} = \text{diag}(9, 11, 11, 22, 14, 0, 23, 25) \text{ meV}.$$
 
The Hamiltonian $h_{\text{t},\tau}(\mathbf{k})$ describes the valley-dependent electronic structure of ABCB within the $\mathbf{k}\cdot\mathbf{p}$ model.
It is structured as a $4 \times 4$ block matrix, where each block is a $2 \times 2$ matrix corresponding to the sublattice degree of freedom. 
Specifically, $h_{\text{t},\tau}(\mathbf{k})$ is given by:
\begin{equation}
h_{\text{t},\tau}(\mathbf{k}) = \begin{pmatrix}
	H_g & M_{AB}^\dagger & M_{ABC}^\dagger & \mathbf{0} \\
	M_{AB} & H_g & M_{AB}^\dagger & M_{ABA}^\dagger \\
	M_{ABC} & M_{AB} & H_g & M_{AB}^\dagger \\
	\mathbf{0} & M_{ABA} & M_{AB} & H_g
\end{pmatrix},
\end{equation} 
The intralayer Hamiltonian $H_g$ for a single graphene layer is defined as:
$$ H_g = \begin{pmatrix} 0 & v\pi_\tau^\dagger \\ v\pi_\tau & 0 \end{pmatrix}, $$
where $v$ represents the intralayer Fermi velocity. $\pi_\tau\equiv\tau p_x + i p_y$, with $(p_x, p_y)$ denoting the momentum around the $K_\tau$ (Dirac) point of the Brillouin zone.

The interlayer coupling blocks, $M_{AB}$, $M_{ABA}$, and $M_{ABC}$ are defined as follows:
$$	M_{AB} = \begin{pmatrix} -v_4\pi_\tau & \gamma_1 \\ -v_3\pi_\tau^\dagger & -v_4\pi_\tau \end{pmatrix}, $$
$$	M_{ABA} = \begin{pmatrix} \frac{\gamma_5}{2} & 0 \\ 0 & \frac{\gamma_2}{2} \end{pmatrix},
M_{ABC} = \begin{pmatrix} 0 & 0 \\ \frac{\gamma_2}{2} & 0 \end{pmatrix}, $$
Here, $\gamma_i$ are the Slonczewski-Weiss-McClure parameters with velocities defined as $v_i = \sqrt{3}a\gamma_i/(2\hbar)$. The hopping parameters are set as $(\gamma_{1},\gamma_{2},\gamma_{5})=(240,-12,40)$ meV and $(\gamma_{0},\gamma_{3},\gamma_{4})=(3560,350,100)$ meV, consistent with our DFT results (Appendix \ref{APP_A}).

Next, we implement the dominant proximity-induced Ising-type SOC described by $\mathcal{H}_{\text{soc}}$. This selectively modifies states associated with the bottom layer. It primarily affects the non-dimer site (blue sites, A$_1$ in Fig.~\ref{fig:1}g) at low-energy scale. $\mathcal{H}_{\text{soc}}$ takes the following form:
\begin{equation}
	\mathcal{H}_{\text{soc}} = \lambda \sum_{\mathbf{k}, \tau = \pm 1} \tau \left( \hat{n}_{\alpha L, \mathbf{k}, \tau, \uparrow} - \hat{n}_{\alpha L, \mathbf{k}, \tau, \downarrow} \right).
\end{equation}

Here, $\alpha$ and $L$ denote sublattice and layer index ($\alpha=A$, $L=1$ in this case). The term $(\hat{n}_{\uparrow} - \hat{n}_{\downarrow})$ denotes the local spin polarization along the out-of-plane (z-axis) direction.
In the low-energy limit, this term reduces to an effective form, $\mathcal{H}_{\text{Ising}} =\lambda\tau_z s_z \hat{P}_{A_1}$, with $\hat{P}_{A_1}$ being the projector onto the A$_1$ sublattice.
Therefore, it acts as an effective valley-dependent, z-axis Zeeman-like field.
As $\mathcal{H}_{\text{soc}}$ is constructed from operators that depend only on $s_z$, it preserves the $U(1)$ spin rotational symmetry along z-axis.
Consequently, the total out-of-plane spin remains a good quantum number.

Rashba SOC is inherently off-diagonal in the sublattice basis, primarily mediating inter-sublattice interactions.
However, low-energy electronic states in ABCB exhibit pronounced sublattice or layer polarization.
This polarization substantially diminishes the efficacy of Rashba SOC\cite{wang_electrical_2024,zollner_proximity_2022,zhumagulov_emergent_2024,gmitra_graphene_2015,liu_layer-dependent_2025,koh_symmetry-broken_2024,koh_correlated_2024}.  
Thus, its contribution is considered negligible in our low-energy effective model. 
 
To describe correlation effects within the system, a local interaction Hamiltonian was constructed (see Appendix \ref{APP_B} for details)\cite{liu_layer-dependent_2025,you_kohn-luttinger_2022,lu_correlated_2022,chatterjee_inter-valley_2022,zhumagulov_emergent_2024,zhumagulov_swapping_2024}. 
\begin{equation}
\begin{aligned}
	\mathcal{H}_{\text{int}} = \int d^2\mathbf{r} \sum_{i} \left( \frac{U}{2} (\hat{n}_{i})^2 
	+ V \hat{n}_{i,K} \hat{n}_{i,K'} - J_H \hat{\mathbf{S}}_{i,K} \cdot \hat{\mathbf{S}}_{i,K'} \right).
\end{aligned}
\end{equation}
Here, the indice $i$ run over all unit cells and sublattices, respectively. We can deconstruct $\mathcal{H}_{\text{int}}$ into three components:
An on-site Hubbard term introduces a Coulomb penalty for double electron occupancy.
A second term describes the density-density interaction between the $K$ and $K^{\prime}$ valleys.
The third term is the Hund's exchange coupling, which for $J_H > 0$ ($J_H < 0$) drives ferromagnetic (antiferromagnetic) alignment between valley-spins.
This exchange interaction is responsible for breaking the independent $SU(2)_K \times SU(2)_{K'}$ spin symmetry of the individual valleys.
As a result, the symmetry is reduced to a single global $SU(2)$ group, corresponding to the conservation of the total spin. 

For our main analysis, we adopt the representative parameterization $V = -U/4$ and $J_H = U/4$, motivated by the common origin of on-site interactions. To confirm our conclusions are not an artifact of this specific ratio, we have investigated the system's behavior across a range of non-local interaction strengths.
As demonstrated in Appendix \ref{APP_D}, varying the ratios of $V/U$ and $J_H$(e.g., from 1/3 to 1/5) slightly shifts the phase boundaries but does not alter the key physical picture.

\subsection{Physical Picture}  
To provide an intuitive physical picture of the pathway toward a topological state, we begin this section with a schematic model. We will first illustrate the established mechanism for the QAH effect in the symmetric ABCA, which is driven by the interplay of interactions, SOC, and an external electric field. Subsequently, we will contrast this with the case of ABCB, elucidating the mechanism for a possible intrinsic QAH state that leverages the system's inherent spontaneous polarization and SOC, even in the absence of an electric field.

\begin{figure}[htb]  
	\centering
	\includegraphics[width=0.5 \textwidth, trim=30 20 0 20, clip]{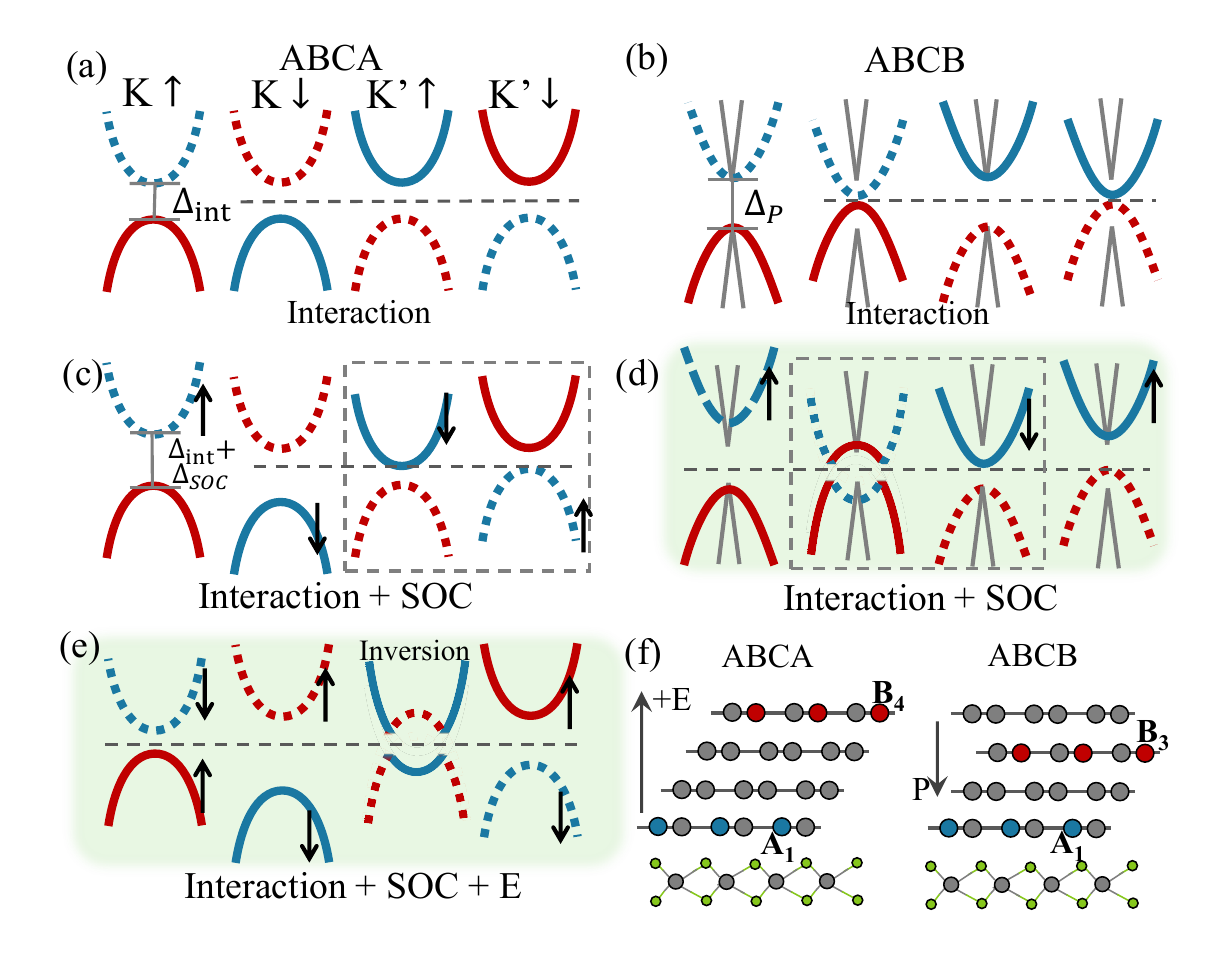}    
	\caption{\label{fig:1} 
		Schematic comparison of interaction-driven symmetry-broken states in ABCA and ABCB.
		(a, c, e) The four spin-valley channels in ABCA under interaction (a), interaction with SOC (c) and with electric field $E$ (e).
		(b, d) The channels in ABCB under interaction (b) and with SOC (d).
		(f) Side view of the lattices defining the non-dimer sites (A$_1$, B$_4$, B$_3$), the direction of $E>0$, and the spontaneous polarization $P$.
		Blue/red bands denote wavefunction localization on A$_1$/B$_4$ (for ABCA) and A$_1$/B$_3$ (for ABCB) sites.
		Solid/dashed bands represent Chern numbers $C=\pm 2$ for ABCA and $C=\pm 3/2$ for ABCB, respectively.
		Black arrows indicate energy shifts, and the dashed box highlights a channel tending towards band inversion.}
\end{figure}  
 
Figure \ref{fig:1} shows the band evolution for ABCA and ABCB under $\mathcal{H}_{\text{int}}$, $\mathcal{H}_{\text{SOC}}$, and electric field. 
We first address ABCA (Fig. \ref{fig:1}a, 1c and 1e).
Possessing inversion symmetry, ABCA is gapless in its non-interacting picture $\mathcal{H}_{\text{eff}}$.
The exchange component of the on-site interaction $\mathcal{H}_{\text{int}}$ drives a spontaneous symmetry breaking to lower the system's total energy. This stabilizes a LAF ground state, where electrons of opposite spins segregate onto opposite layers to maximize the favorable exchange energy.
This emergent, spin-dependent ordering effectively localizes electrons of a given spin onto a single external layer (A$_1$ or B$_4$) leading to a correlation gap (Fig.~\ref{fig:1}a).
Without loss of generality, we consider the state where occupied spin-$\uparrow$ (spin-$\downarrow$) channels from both valleys reside on the top (bottom, WSe$_2$-proximal) layer.

The SOC subsequently modifies the orbital energies on the WSe$_2$-proximal bottom layer, breaking the four-fold spin-valley degeneracy that exists with interactions alone. As depicted in Fig.~\ref{fig:1}c, this is achieved by shifting the energy of the unoccupied states (blue bands) in a spin-valley locked manner: the $K\uparrow$ and $K'\downarrow$ channels are raised in energy, while the $K\downarrow$ and $K'\uparrow$ channels are lowered, creating the possibility of a single-channel band inversion without an electric field. This splitting pushes the system towards a topological instability, where a band inversion in one of the lowered-energy channels (highlighted in the dashed box) can readily induce a quantum spin Hall (QSH) phase.

Building on the SOC-modified state (Fig. \ref{fig:1}c), a perpendicular upward electric field is applied. This field places the bottom-most layer at a lower potential energy and the top-most layer at a higher one. Consequently, across all four channels, the bands localized on the bottom A$_1$ layer (blue bands) universally down-shift in energy, while those on the top B$_4$ layer (red bands) up-shift. Due to the pre-existing energy splitting from SOC, this differential shift selectively drives only the $K'\uparrow$ channel to invert at a critical field $E_c$. Since the low-energy bands of N-layer RG carry a Chern number with magnitude $N/2$, this single inversion in the $N=4$ system flips its contribution from $+2$ to $-2$, resulting in the total $C=4$ QAH state depicted in Fig. \ref{fig:1}e.  

ABCB differs significantly (Fig. \ref{fig:1}b and \ref{fig:1}d).
It intrinsically lacks inversion or mirror symmetry, resulting in a gapped non-interacting band structure.
This gap, arising from the material's inherent polarization, is analogous to the gap opened by an external perpendicular electric field in a system without intrinsic polarization.
For its cubic bands, occupied states are localized on the B$_3$ layer, and unoccupied states on the WSe$_2$-proximal A$_1$ layer. The electron interactions favor a LAF ground state. This is driven by an effective mean-field potential, the `LAF exchange field', which breaks the system's spin degeneracy. This symmetry breaking creates two possible states: one where the gap of the spin-up channels ($K\uparrow, K'\uparrow$) shrinks, and another where the gap of the spin-down channels ($K\downarrow, K'\downarrow$) shrinks. 
In the state shown (Fig. \ref{fig:1}b), the latter case is realized: the gap for the spin-down channels decreases while the gap for the spin-up channels increases, though the system remains a fully polarized insulator.

Starting from the interaction-driven state (Fig. \ref{fig:1}b), the subsequent application of SOC acts on the bands associated with the A$_1$ layer (blue bands). This introduces a spin-valley locked energy shift: the A$_1$ bands for the $K\uparrow$ and $K'\downarrow$ channels are raised in energy, while those for the $K\downarrow$ and $K'\uparrow$ channels are lowered. As illustrated in Fig. \ref{fig:1}d, this differential shift selectively shrinks the energy gap for the $K\downarrow$ and $K'\uparrow$ channels (highlighted in the dashed box) making a band inversion favorable. The inversion of a single channel flips its contribution from $C=+3/2$ to $C=-3/2$, a value stemming from the 3RG fragment, yielding a QAH state with a net Chern number of $C=3$. 

In summary, the two cases follow distinct topological pathways. For symmetric ABCA, interactions induce a topologically trivial ($C=0$) LAF gap, necessitating an electric field to overcome this gap and invert selective channels. Conversely, for the intrinsically polarized ABCB, the interaction acts on an existing layer-polarized insulator (LPI) gap. The drive towards the energetically favorable LAF order inherently shrinks the gap for two spin-valley channels, an effect that substitutes for an electric field and provides a natural pathway toward a field-free QAH state.

To provide a clear physical intuition for how intrinsic polarization substitutes for an external electric field in driving the QAH state, we describe the mechanisms in ABCB and ABCA from a different perspective. The crucial first step in both systems is establishing a gapped, LPI state, which serves as the foundation for subsequent topological phase induction by electron interactions and SOC. In the single-particle picture, the gapless ABCA requires an external electric field to create this initial LPI state. Conversely, in ABCB, the intrinsic spontaneous polarization inherently provides this LPI state without external stimulus. Thus, ABCB's built-in field creates the gapped, polarized state, just like ABCA's external field. Upon this LPI state, the interplay of interactions and SOC then breaks remaining degeneracies and inverts specific band gaps to yield a net Chern number.

Having established the physical pictures for achieving the QAH state in both stacking configurations, we now further discuss how the emergence of these topological phases is governed by the interaction strength U. For ABCA, if SOC dominates a weak interaction, it would simultaneously invert the K↓ and K'↑ channels. Since all four occupied bands would carry a Chern number of +2, this would lead to a C=8 state. For ABCB, if the interaction is weak, it leaves the intrinsic gap of the K↓ and K'↓ channels too large for SOC to induce a topological transition without an external field. Conversely, an excessively strong interaction inverts both channels into a trivial LAF state. From this state, a zero-field QAH phase can still emerge if the inverted gap is smaller than the SOC strength, allowing SOC to induce a second inversion in the K'↓ channel. If the inverted gap is larger than the SOC strength, an external field again becomes necessary.

\subsection{Electronic Properties without Interaction} 
\label{sectionc}
\begin{figure}[htb]
	\centering
	\includegraphics[width=0.5 \textwidth, trim=20 60 20 30,clip]{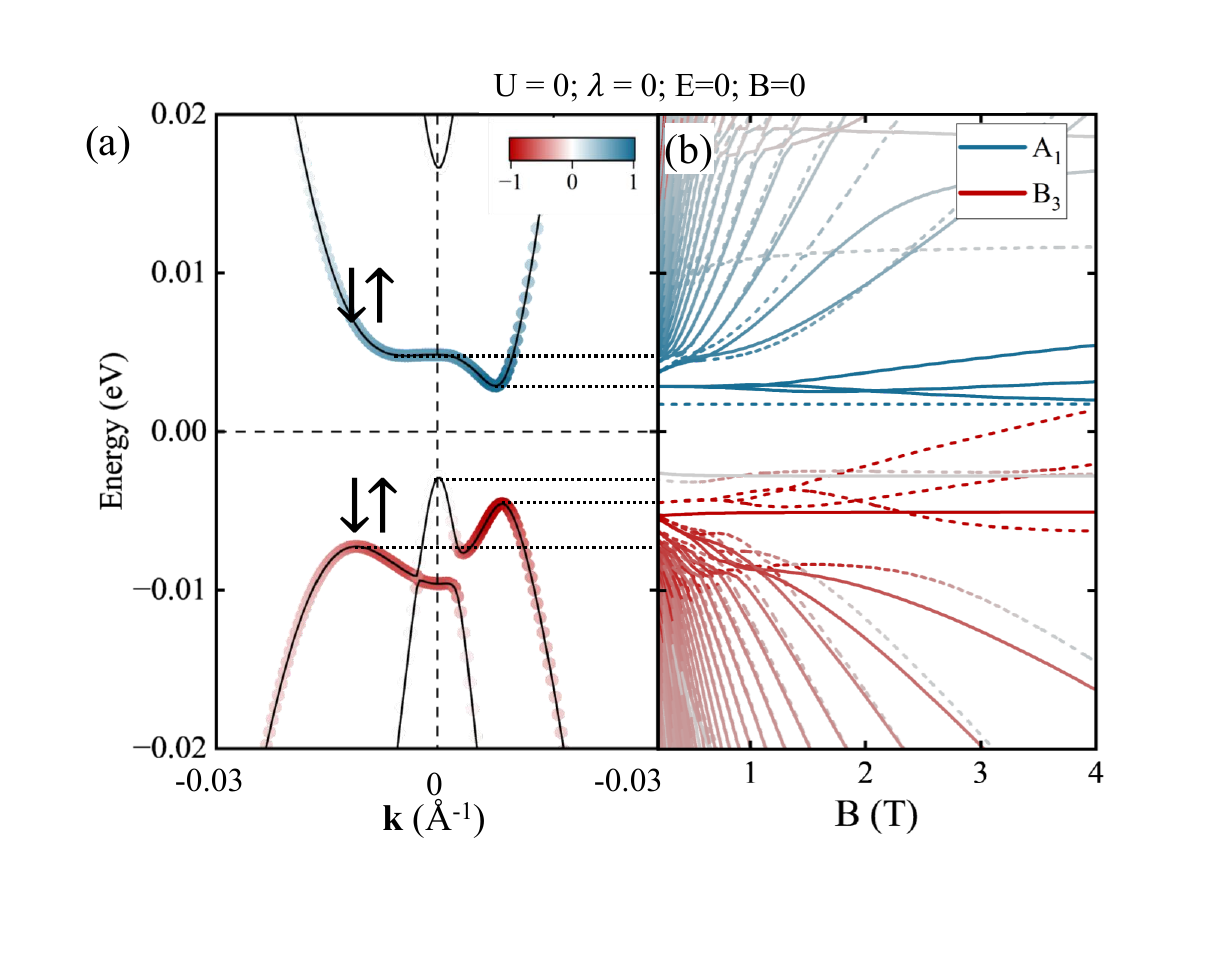} 
	\caption{\label{fig:2} Bands and Landau levels of ABCB for the non-interacting Hamiltonian $\mathcal{H}_{\text{eff}}$ without electric field and SOC. (a) Site-resolved projection of the band structure with red representing the B$_3$ site (on layer 3) and blue representing the A$_1$ site (on layer 1). Black arrows indicate the spin direction of the bands. (b) Calculated Landau levels as a function of magnetic field. Solid (dashed) lines represent states from one valley (the other valley). }  
\end{figure}

To validate the physical picture proposed in the preceding section for a potential intrinsic QAH state in ABCB, we now present a systematic analysis based on our model Hamiltonian. We will build up the full picture step-by-step: in this Section \ref{sectionc}, we first characterize the electronic structure in the non-interacting limit to establish a baseline. Subsequently, we will investigate the influence of interactions alone (Section \ref{sectiond}), and finally, we will demonstrate how interactions and SOC act in concert to realize the QAH state (Section \ref{sectione}).

Based on our fitting results, the on-layer potential energies for the first (bottom) to fourth (top) layers are 20, 33, 14 and 48 meV, respectively. This asymmetric potential profile generates a net, downward-oriented spontaneous polarization, which induces an intrinsic indirect band gap of approximately 5.7 meV in the ground state of ABCB, as depicted in Fig. \ref{fig:2}a.

The low-energy electronic structure comprises four bands: two with linear dispersion and two exhibiting quasi-cubic dispersion.
The linearly dispersing bands are predominantly contributed by the layer 2 and layer 4. 
Furthermore, the DOS for linear bands near the Fermi energy is substantially lower than that of the quasi-cubic bands, implying weak electron correlation effects.
Additionally, these linear bands possess negligible Berry curvature, suggesting their contribution to the topological character of the occupied states is likely limited.

For the quasi-cubic bands in a fully polarized state, electrons are predominantly localized on the A$_1$ site (on layer 1), while holes reside mainly on the B$_3$ site (on layer 3), leading to a LPI state.
Each of these quasi-cubic bands is four-fold degenerate at zero field. A small associated energy gap of approximately 8 meV facilitates the subsequent emergence of symmetry-broken states.
Moreover, under a magnetic field, this fully polarized state leads to Landau levels with broken valley degeneracy but preserved spin degeneracy, as Zeeman splitting is considered negligible.
Consequently, the degeneracy of certain states is reduced from four-fold to two-fold, as illustrated in Fig. \ref{fig:2}b.

\begin{figure}[htb]
	\centering
	\includegraphics[width=0.5 \textwidth, trim=60 30 20 50,clip]{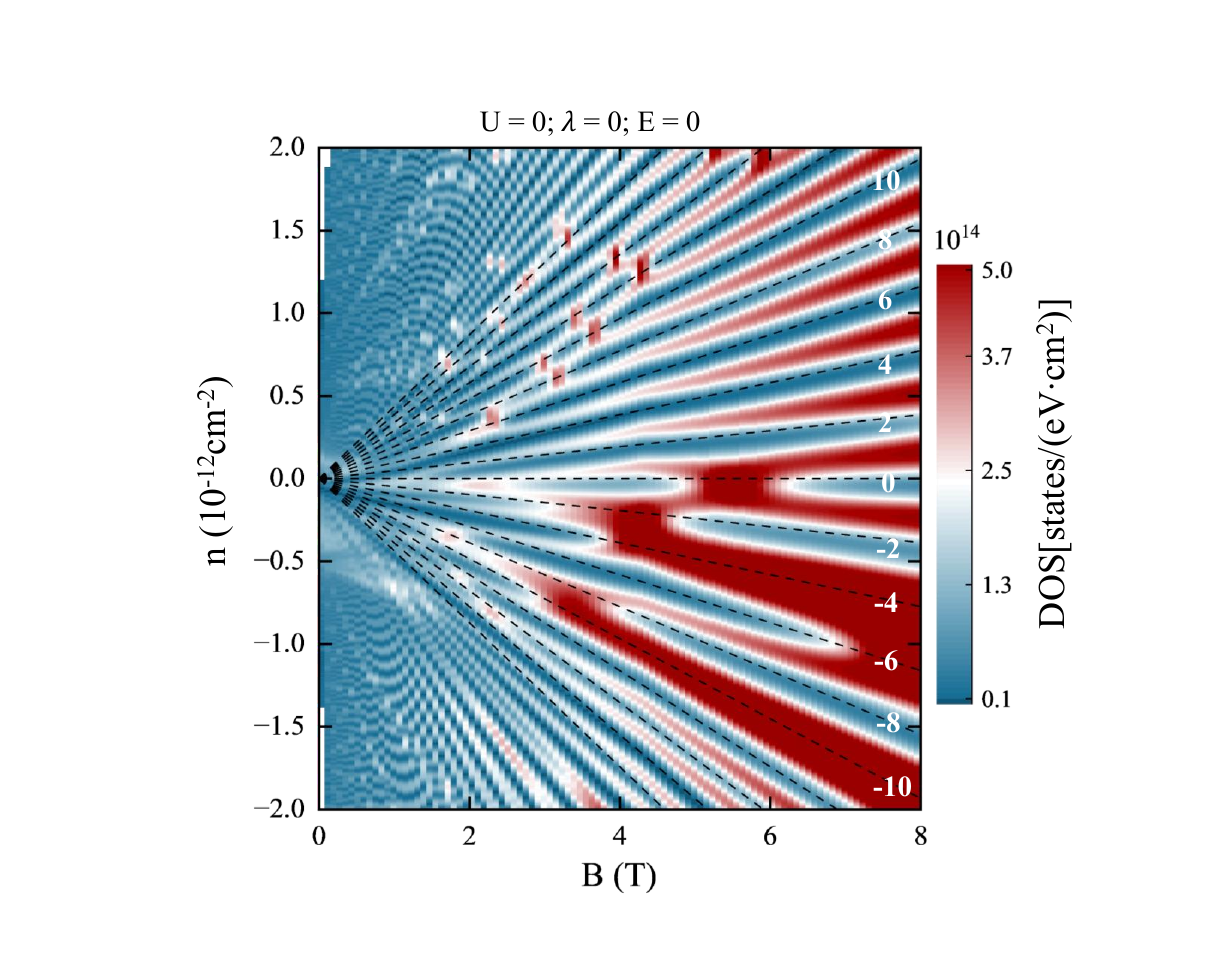}
	\caption{\label{fig:3} Landau fan diagram for ABCB. The plot shows the DOS as a function of carrier density $n$ (negative values denote hole doping, positive values denote electron doping) and magnetic field $B$. Red (blue) color indicates high (low) DOS.}
\end{figure} 
Figure \ref{fig:3} shows the calculated Landau fan diagram (DOS vs. $n$ and B) for ABCB, revealing signatures of the quantum Hall effect.
Away from the CNP, higher-order Landau levels form a series of clear fan stripes, which appear at a sequence of filling factors $\nu = \pm 2, \pm 4, \dots$. The constant spacing of $\Delta\nu=2$ is an indication of a partially reduced degeneracy. Specifically, while the spin degeneracy ($g_s=2$) remains intact, the intrinsic polarization of the ABCB lattice breaks the valley degeneracy, thus halving the original four-fold degeneracy and producing the observed sequence.

\begin{figure}[htb]  
	\centering
	\includegraphics[width=0.45 \textwidth, trim=40 50 30 40,clip]{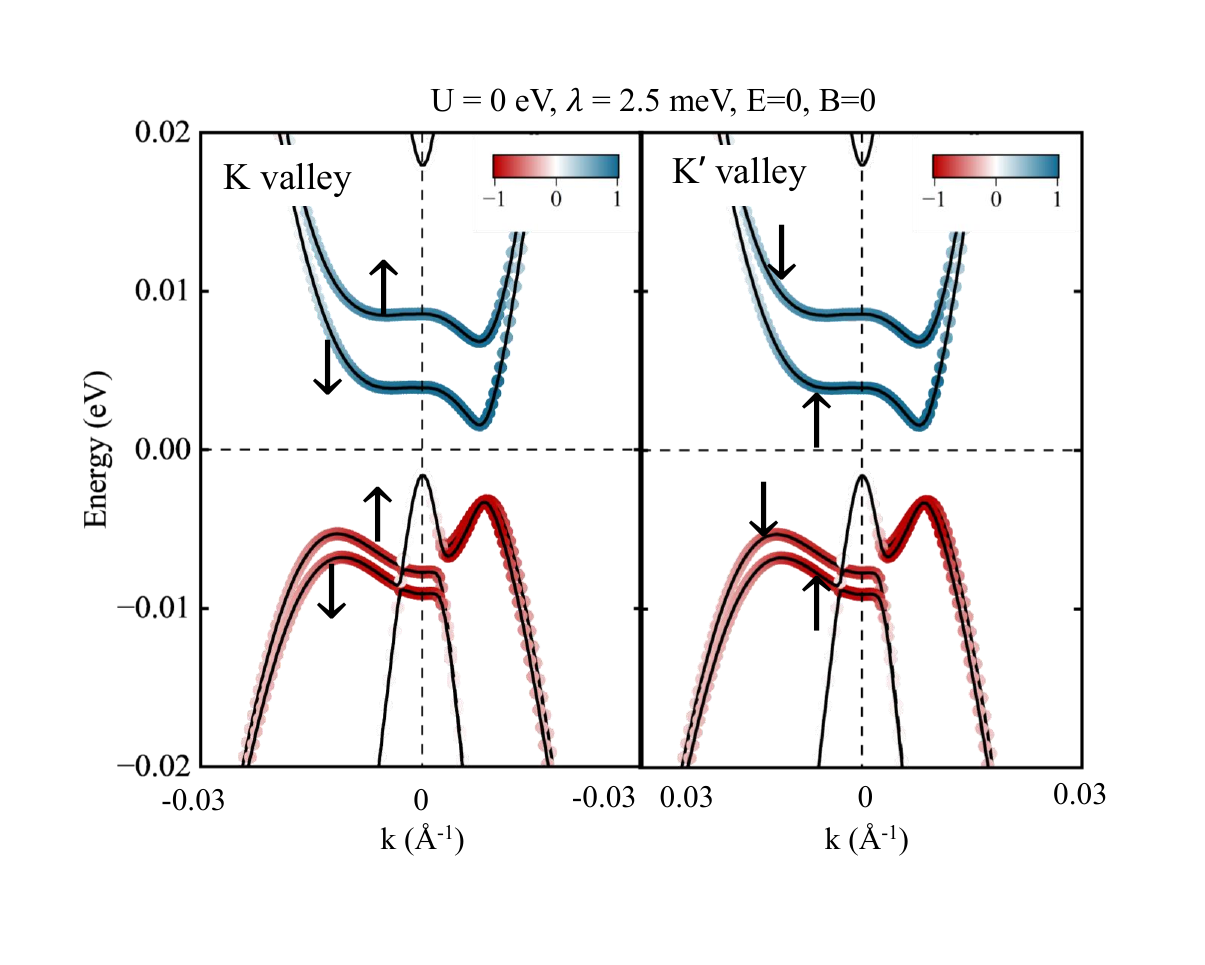}    
	\caption{\label{fig:4}
		Projected band structure for ABCB calculated with an SOC strength of $\lambda = 2.5 \text{ meV}$.
		The left and right panels display results for the $K$ and $K^{\prime}$ valleys, respectively. Black arrows indicate the spin polarization of the bands. Red (blue) representing the B$_3$ (A$_1$) site.}  
\end{figure}

Upon introducing SOC with a strength of $\lambda=2.5 \text{ meV}$\cite{zihlmann_large_2018,gmitra_proximity_2017} (Fig. \ref{fig:4}), the four-fold degeneracy of the quasi-cubic conduction bands is reduced. In this low-energy picture, the SOC primarily acts on the A$_1$ site, where these unoccupied bands are localized. This introduces a spin-valley locked energy shift as discussed above. As the valence bands on the B$_3$ site remain largely unperturbed, this downward shift of the $K\downarrow$ and $K^{\prime}\uparrow$ conduction bands selectively shrinks the band gap. This behavior signals a tendency towards a topological phase transition, where a sufficiently large $\lambda$ is expected to induce a full band inversion and drive the system into a QSH phase.

\subsection{Coupling between Interaction and Polarization}
\label{sectiond}
For comparison, we first consider ABCA. Its low-energy bands feature a quartic dispersion, giving rise to a high DOS near the Fermi level. Consequently, strong correlations at low temperatures drive a phase transition from metallic to a LAF ground state.
Self-consistent calculations confirm this LAF configuration as the ground state, proving energetically favorable over other competing phases—such as the QAH, QSH, and LPI states—irrespective of the initial trial state\cite{liu_spontaneous_2024}.

\begin{figure}[htb] 
	\centering
	\includegraphics[width=0.45 \textwidth, trim=40 50 30 40,clip]{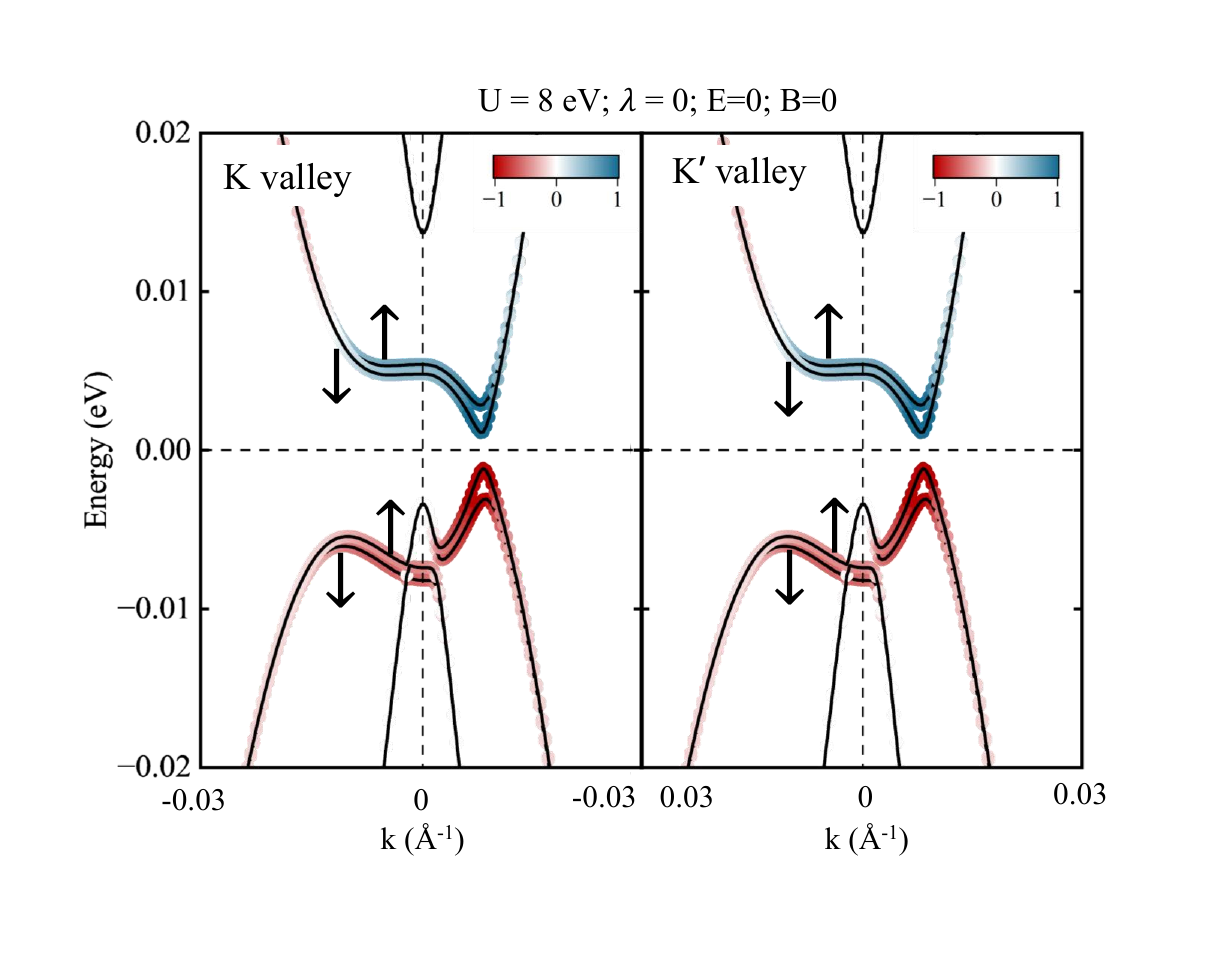}    
	\caption{\label{fig:5}
		Projected band structure for ABCB calculated with an interaction strength of $U = 8 \text{ eV}$.	The left and right panels display results for the $K$ and $K^{\prime}$ valleys, respectively. Black arrows indicate the spin polarization of the bands. Red (blue) representing the B$_3$ (A$_1$) site.
		}
\end{figure}

ABCB displays analogous physical properties to ABCA: its intrinsic band structure also features a Lifshitz transition near the CNP, and its maximum DOS is quantitatively comparable\cite{fischer_spin_2024, unpublished}. 
However, their response to interactions is markedly different. 
The symmetric and initially gapless ABCA readily transitions into a LAF state, as an interaction of $U=8 \text{ eV}$ is sufficient to open an interaction gap. 
In contrast, the ABCB is intrinsically a gapped LPI. 
To become a LAF state, the interaction must be strong enough to first close the initial cubic-bands gap for a specific spin channel. 
Consequently, an interaction strength of $U=8 \text{ eV}$, while sufficient for ABCA, is insufficient to drive this transition in ABCB, which remains in the LPI state (Fig. \ref{fig:5}).

Therefore, in ABCB, the spontaneous polarization opens an intrinsic gap that suppresses the formation of an interaction-driven LAF state. 
At an interaction strength of $U=8 \text{ eV}$, this suppression leaves a small residual gap of only $\sim 2 \text{ meV}$, which can potentially be overcome by SOC without an external electric field. 
Paradoxically, while this mechanism is absent in the gapless ABCA, it is precisely this suppression of the LAF phase that enables the field-free QAH state in ABCB.
In ABCA, interactions first open a much larger gap of approximately $15 \text{ meV}$ for $U=6 \text{ eV}$, which then requires a substantial electric field to close.

\subsection{QAH State with Interaction and SOC}
\label{sectione}
This section details the emergence of the QAH state. We will address three key aspects. First, we demonstrate that at strong interaction strengths, the system spontaneously enters an intrinsic QAH state, even at zero electric field. Second, for the case of weaker interactions, we show that a finite perpendicular electric field is required to trigger the topological phase and we determine the specific electric field window for its realization. Finally, to provide a comprehensive picture of this tunability, we present a phase diagram of the Chern number as a function of interaction strength, SOC, and the applied electric field.

As discussed above, the potentially intrinsic and topologically non-trivial states of ABCB arise from the interplay between SOC and interactions $U$, which must act in concert to overcome a non-interacting gap of $\sim 8 \text{ meV}$ separating the occupied and unoccupied cubic bands in a specific channel.
As $U$ and SOC increase, they collaboratively reduce this gap, making it possible for a critical combination of the two to induce a topological transition without any electric field. 

\begin{figure}[htb]  
	\centering 
	\includegraphics[width=0.45 \textwidth, trim=40 50 30 40,clip]{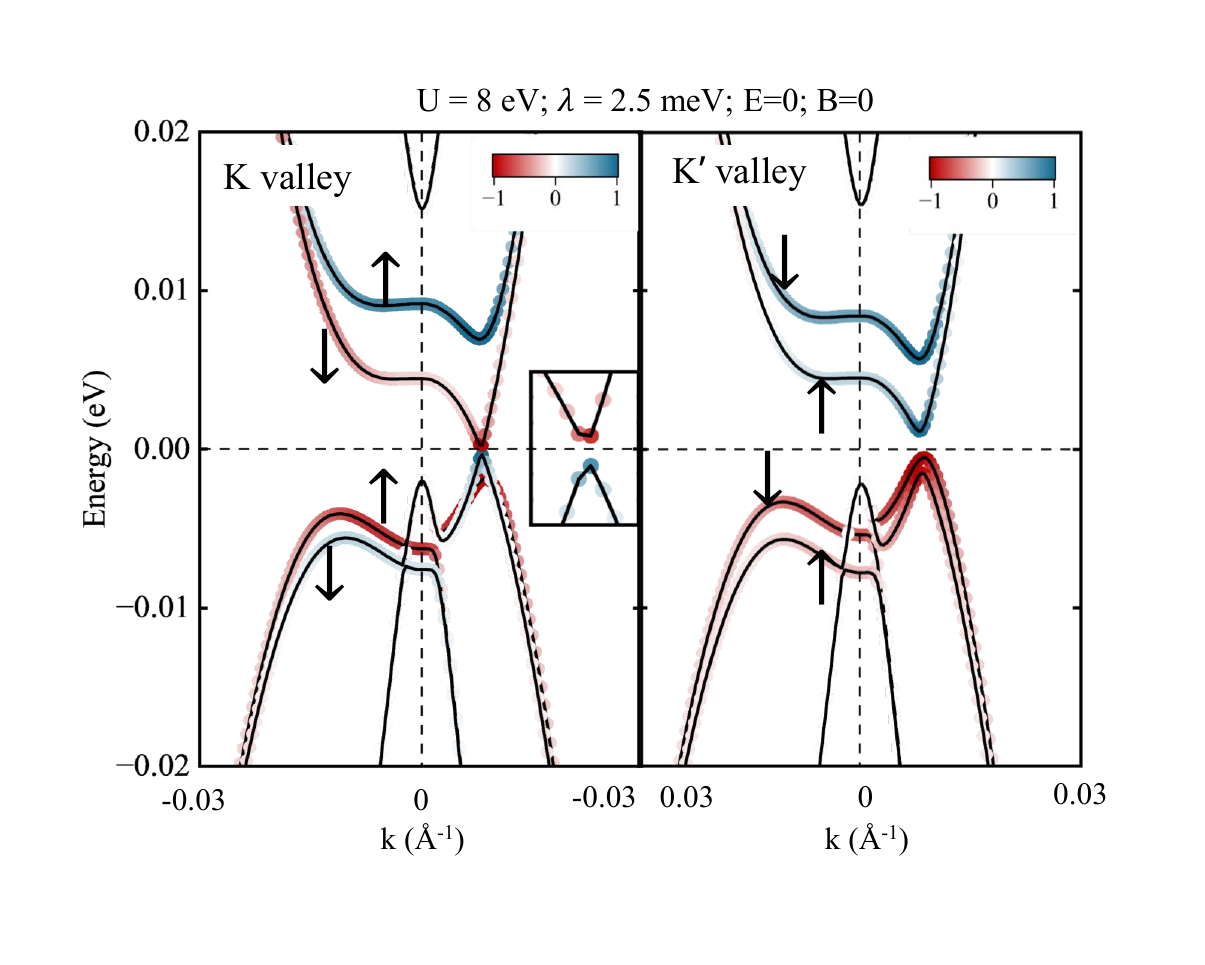}   
	\caption{\label{fig:6}
		Projected band structure for ABCB calculated with an SOC strength of $\lambda = 2.5 \text{ meV}$ and interaction strength of $U = 8 \text{ eV}$. The left and right panels display results for the $K$ and $K^{\prime}$ valleys, respectively. Black arrows indicate the spin polarization of the bands. Red (blue) representing the B$_3$ (A$_1$) site.}    
\end{figure}

The SOC effect tends to invert the $K\downarrow$ and $K^{\prime}\uparrow$ channels while interactions favor inversion of the $K\downarrow$ and $K^{\prime}\downarrow$ channels, allows for a scenario where only the $K\downarrow$ channel inverts. 
Indeed, as shown in Fig. \ref{fig:6}, for strong interactions ($U=8 \text{ eV}$) and SOC ($\lambda=2.5 \text{ meV}$), the system undergoes a spontaneous band inversion exclusively in the $K\downarrow$ channel. This overcomes the intrinsic gap, resulting in a QAH state with a net Chern number $C=3$ even without an external electric field.

\begin{figure}[htb] 
	\centering
	\includegraphics[width=0.5 \textwidth, trim=10 50 0 50,clip]{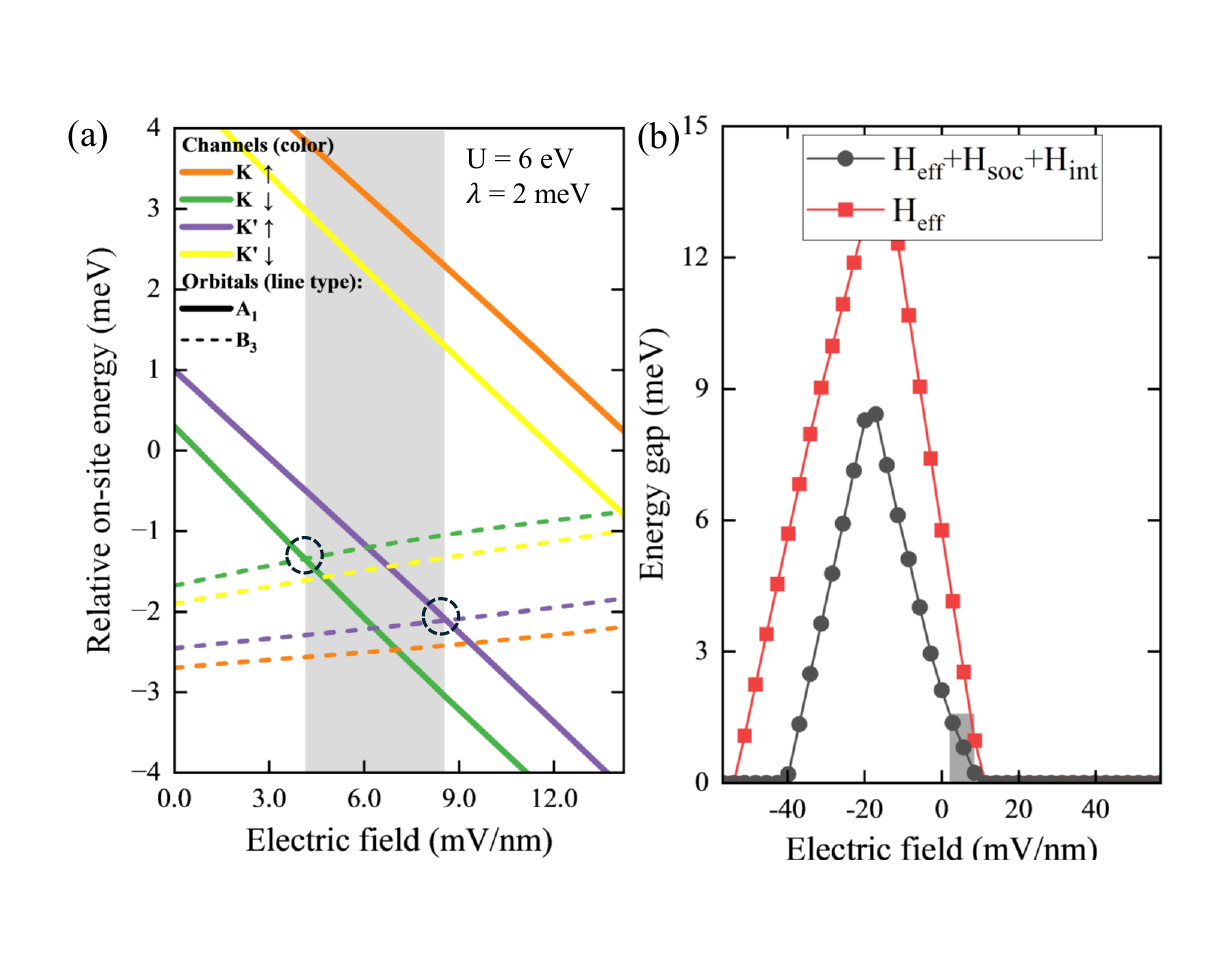}     
	\caption{\label{fig:7}
		(a) Orbital on-site energies for ABCB as a function of the applied electric field.
		The vertical axis represents the relative on-site energies, with the energy origin chosen for clarity.
		The orange, green, purple, and yellow lines represent the $K\uparrow$, $K\downarrow$, $K'\uparrow$, and $K'\downarrow$ channels, respectively.
		Solid (dashed) lines indicate orbitals on the non-dimer sites of the A$_1$(B$_3$).
		The grey shaded area indicates the electric field regime corresponding to the QAH states.
		(b) Band gap as a function of the electric field.
		The red and black lines denote gaps derived from $\mathcal{H}_{\text{eff}}$ (e xcluding SOC and interactions) and the $\mathcal{H}_{\text{tot}}$, respectively.}    
\end{figure}

Weaker interaction of $U=6 \text{ eV}$—a value consistent with experimental fits for the ABCA (see Appendix \ref{APP_F} for details)—is insufficient to induce this band inversion. However, the topological transition can still be triggered by applying a modest, upward-oriented field ($E \in [4, 8.5]$~mV/nm, Fig. \ref{fig:7}a), which facilitates the complete band inversion necessary for the QAH state.

This naturally leads to the question of what range of electric field is required to trigger the QAH phase under the weaker interactions ($U=6 \text{ eV}$). To determine this, we calculated the on-site energies for the orbitals at sites $\text{A}_1$ and $\text{B}_3$—which respectively contribute to the unoccupied and occupied states of the cubic bands—as presented in Fig. \ref{fig:7}a.
For an upward-oriented electric field, the self-consistent redistribution of charge across the lattice sites creates a screening effect. This screening is more pronounced at the interior site B$_3$ compared to the bottom-most site A$_1$, resulting in the on-site energy of A$_1$ being more substantially influenced by the applied electric field.

Initially, in the absence of an electric field, the on-site energy associated with site $\text{B}_3$ is lower for all channels.
Subsequently, the system transitions into a topologically non-trivial phase at an electric field of $E=4 \text{ mV/nm}$, triggered by a band inversion in the $K\downarrow$ channel (left dashed circle in Fig.~\ref{fig:7}a). This phase persists until the field $E$ reaches $8.5 \text{ mV/nm}$, where a second band inversion in the $K^{\prime}\uparrow$ channel (right dashed circle) returns the total Chern number to zero, causing the system to exit the topological state.
Thus, for interaction strength of $U=6 \text{ eV}$, the electric field window for realizing the QAH state spans from $4 \text{ mV/nm}$ to $8.5 \text{ mV/nm}$, a range somewhat narrower than that experimentally observed for ABCA (Appendix \ref{APP_F}).
Furthermore, Fig. \ref{fig:7}b illustrates the evolution of the band gap with the applied electric field. This trend remains largely consistent between the non-interacting and interacting frameworks, with discrepancies primarily in the numerical magnitudes.
This consistency suggests that experimentally discerning the presence of significant interaction effects solely from the dependence of the longitudinal resistance, $R_{xx}$, on the displacement field might be challenging, a situation markedly different from ABCA, where the experimental behavior starkly deviates from non-interacting predictions\cite{liu_spontaneous_2024}. 

The differing response of the band gap to the electric field in Fig.~\ref{fig:7}b also reveals a band-dependent screening mechanism. In the range of $E=11$ to $-17~\text{mV/nm}$, where the gap is formed by strongly correlated quasi-cubic bands, the interacting case ($\mathcal{H}_{\text{eff}}+\mathcal{H}_{\text{soc}}+\mathcal{H}_{\text{int}}$, black curve) exhibits a smaller slope than the non-interacting one ($\mathcal{H}_{\text{eff}}$, red curve). 
This is because the formation of the band gap involves the cubic bands.
Interactions drive a self-consistent charge rearrangement of cubic bands that screens the electric field, thereby weakening its effect on the gap. 
Beyond $E \approx -17~\text{mV/nm}$, the gap is decided only by linear bands. In this regime, the screening becomes negligible, causing the slope of the interacting case to approach that of the non-interacting case. 

\begin{figure}[htb]  
	\centering
	\includegraphics[width=0.55 \textwidth, trim=60 70 0 60,clip]{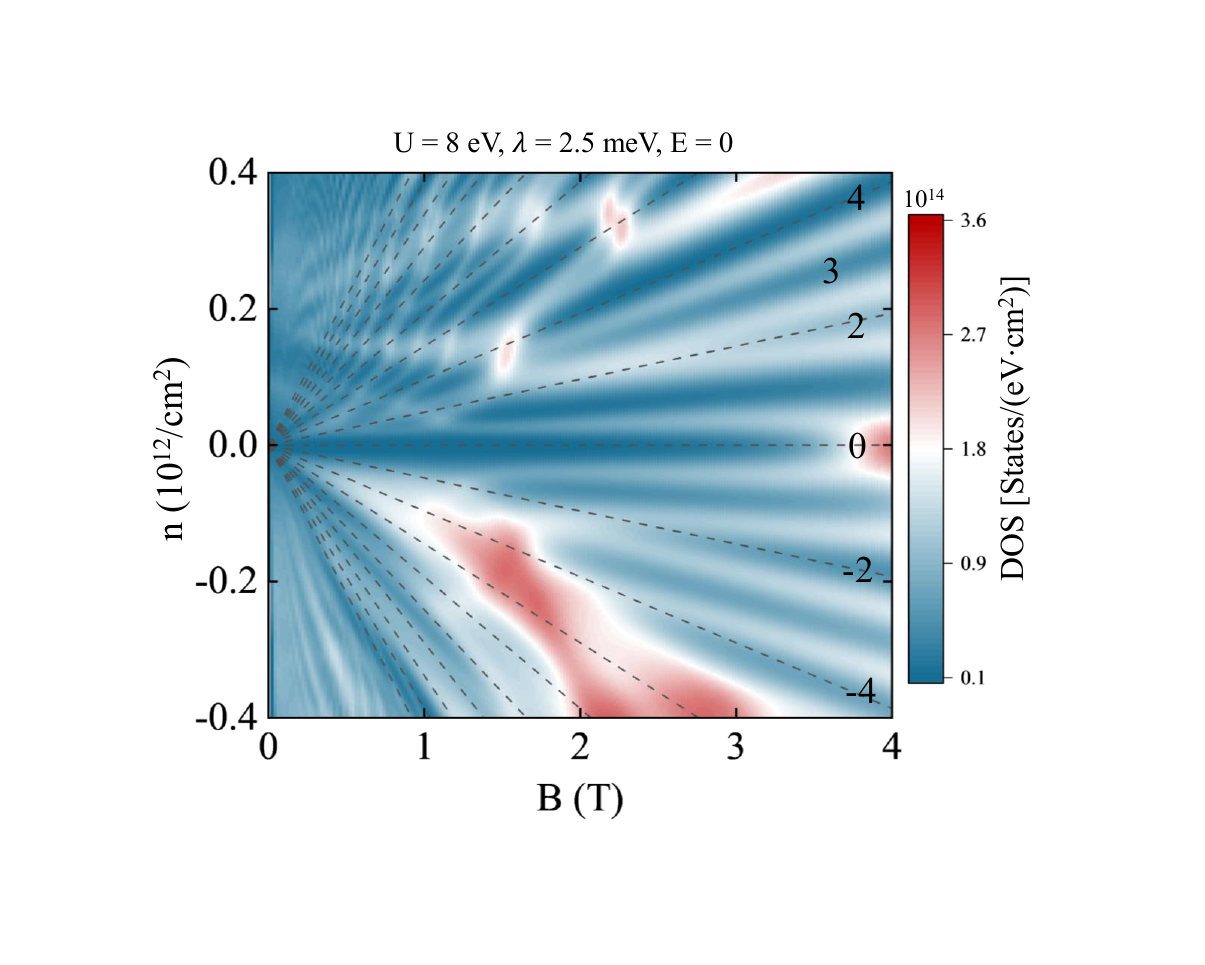} 
	\caption{\label{fig:8}
		DOS for ABCB as a function of $n$ and magnetic field $B$ with $U=8$ eV and $\lambda=2.5$ meV. 
		The numerical labels denote the integer filling factors $\nu$. For clarity, black dashed lines are overlaid to guide the eye along the trajectories of even $\nu$.}
\end{figure}

Having established that the system spontaneously enters a QAH state with a calculated Chern number $C=3$ at zero magnetic field (Fig.~\ref{fig:6}), we now investigate its characteristic response to an applied magnetic field. Figure~\ref{fig:8} presents the calculated Landau fan diagram for this interaction- and SOC-driven topological phase. 
A key feature in the spectrum is the emergence of a new state at the odd integer filling factor $\nu=3$. This state appears alongside the expected sequence of states at even filling factors ($\nu = 0, \pm 2, \pm 4$), which are guided by dashed lines for clarity. The presence of this prominent $\nu=3$ quantum Hall state is a direct consequence of, and consistent with, the zero-field Chern number $C=3$. 

To systematically investigate how interactions $U$, SOC $\lambda$, and electric field $E$ interplay to define the topological phase boundaries, we map out the phase diagram, which reveals that sufficient electronic correlation is a prerequisite for the topological state, as the QAH phase emerges only for an interaction strength of $U > 4~\text{eV}$. 

\begin{figure}[htb] 
	\centering
	\includegraphics[width=0.48 \textwidth, trim=70 20 60 20,clip]{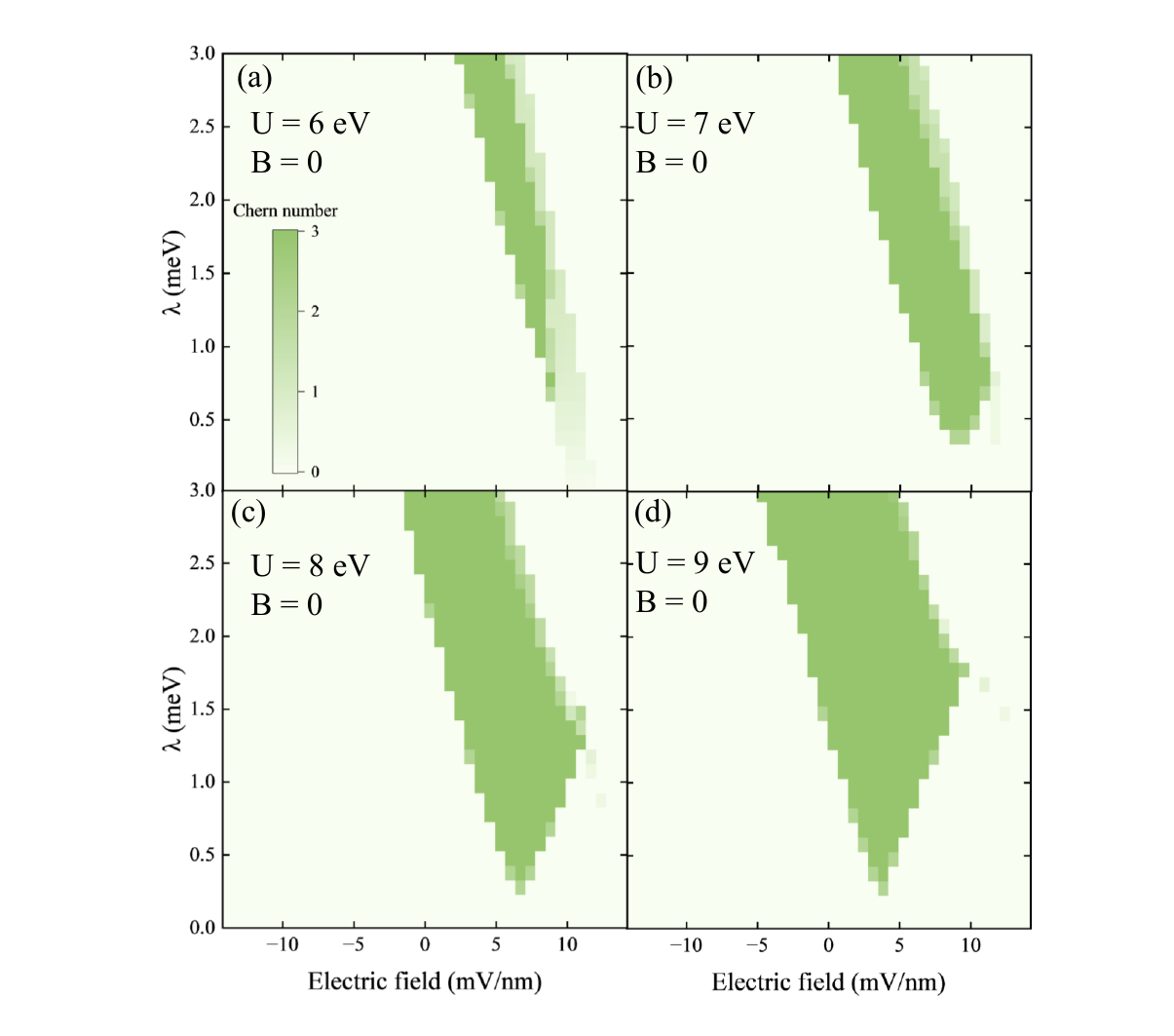}    
	\caption{\label{fig:9}Color map of Chern number for ABCB as a function of $\lambda$ and $E$ for (a) $U=6\text{ eV}$, (b) $U=7\text{ eV}$, (c) $U=8\text{ eV}$ and (d) $U=9\text{ eV}$ .}    
\end{figure}

Figure~\ref{fig:9} shows that the interplay between $U$ and $\lambda$ is crucial for achieving a zero-field QAH state.
While an interaction of $U = 6~\text{eV}$ (Fig.~\ref{fig:9}a) is insufficient to induce a QAH state at $E=0$ for any $\lambda$ up to $3~\text{meV}$, increasing the interaction strength makes it possible. 
At $U=8~\text{eV}$ (Fig.~\ref{fig:9}c), a zero-field QAH state is triggered at $\lambda \approx 2.5~\text{meV}$, and at $U=9~\text{eV}$ (Fig.~\ref{fig:9}d), the required SOC strength is further reduced to $\lambda \approx 1.5~\text{meV}$. This demonstrates that stronger on-site interactions can compensate for a weaker SOC in stabilizing the zero-field topological phase.

To further quantify this tunability, we analyze the required electric field window at a fixed SOC of $\lambda = 2.5~\text{meV}$. As $U$ increases, the window for the QAH phase both shifts towards negative fields and expands in width. For $U=6~\text{eV}$, the window is $3.5$ to $8~\text{mV/nm}$ (a width of $4.5~\text{mV/nm}$). This expands for $U=7~\text{eV}$ ($2.5$ to $8.5~\text{mV/nm}$; width $6~\text{mV/nm}$), $U=8~\text{eV}$ ($0$ to $7.5~\text{mV/nm}$; width $7.5~\text{mV/nm}$), and $U=9~\text{eV}$ ($-2.5$ to $6.5~\text{mV/nm}$; width $9~\text{mV/nm}$). 

Finally, we compare these findings to the ABCA graphene configuration. While the QAH phase window in ABCB is generally narrower than in ABCA, the required electric field is significantly smaller. 
For instance, at $U = 6~\text{eV}$ and $\lambda = 2~\text{meV}$ (see Appendix \ref{APP_F} for details), the QAH phase in ABCA emerges in two side of electric field windows, each $7.5~\text{mV/nm}$ wide, requiring a large field with magnitude $|E|$ between $20$ and $27.5~\text{mV/nm}$.
In contrast, under the same parameters, the window in ABCB is narrower at $3.5~\text{mV/nm}$ ($E=5$ to $8.5~\text{mV/nm}$), but the necessary electric field is substantially lower. 
This highlights a key advantage of the intrinsic polarization in ABCB for realizing topological states with smaller external stimuli.

\subsection{Symmetry Breaking Metal States under Electric Field and Doping}

The preceding analyses reveal that, under the combined SOC and interactions, the degeneracy of nearly all cubic dispersion bands is reduced to its minimum, compared to the non-interacting Hamiltonian $\mathcal{H}_{\text{eff}}$.
This observation naturally suggests that modulating $n$ under these conditions could give rise to intriguing symmetry-broken metallic phases.

Figure \ref{fig:10} maps out the rich phase diagram in the plane of $E$ and $n$ at a fixed $U=6$ eV. 
Notably, at the CNP, sweeping $E$ from $-12$ to $12 \text{ mV/nm}$ induces successive metal-insulator-metal transitions.
At $E=0$, adjustment of $n$ (from $-0.2 \times 10^{12} \text{ cm}^{-2}$ to $0.7 \times 10^{12} \text{ cm}^{-2}$) allows for the ground state to be switched between SP and SVL configurations.
The former is an interaction-induced spin-polarized state where a single spin flavor (spin-$\downarrow$) features bands crossing the Fermi level. The Fermi surface comprises three hole pockets engendered by the trigonal-warping effect, with these holes originating from the B$_3$ site, and the overall conductivity is dominated by contributions from the spin-$\downarrow$ channel.
The latter (SVL) is an SOC-induced spin-valley-locked state, wherein the $K\downarrow$ and $K^{\prime}\uparrow$ channels each possess one band traversing the Fermi energy, with each channel corresponding to an electron pocket derived from the A$_1$ site.

Owing to the substantial DOS generated by the trigonal-warping effect within ABCB, linear bands do not emerge even up to an electron doping $n=0.7 \times 10^{12} \text{ cm}^{-2}$ at $E=0$, and the SVL state maintains its stability.
Subsequently, commencing from the SVL state, the application of an $E=6 \text{ mV/nm}$ coupled with a further increase in electron doping ($n>0$) can induce a transition to a type-I SVP state.
This state is characterized by an additional contribution from the $K^{\prime}\downarrow$ channel superimposed on the SVL baseline. Under these conditions, three channels are conducting while one remains insulating.
\begin{figure}[htb]  
	\centering
	\includegraphics[width=0.46 \textwidth, trim=10 0 0 0,clip]{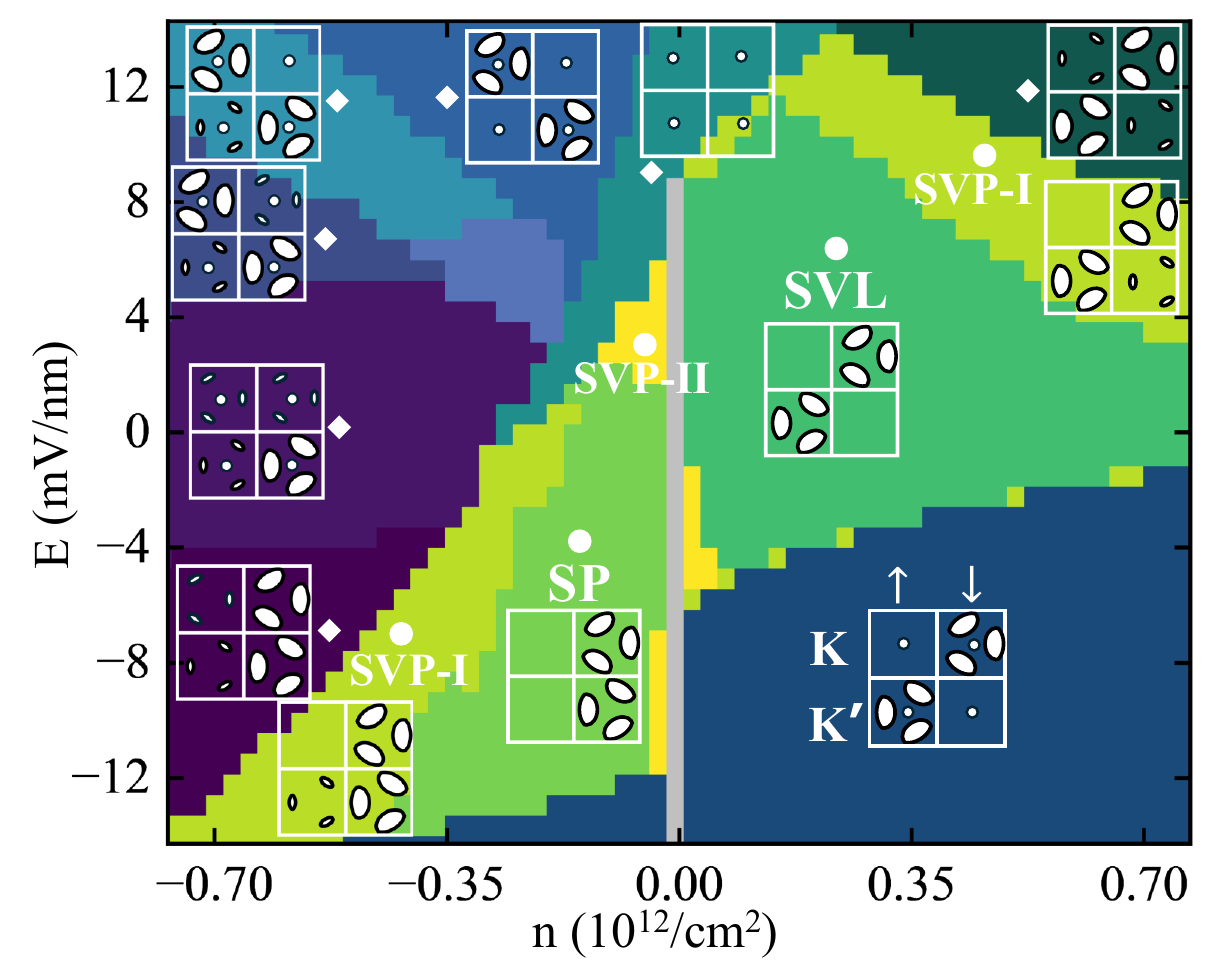}   
	\caption{\label{fig:10}
		Phase diagram for ABCB as a function of $E$ and $n$ for $U=6$ eV, $\lambda=2$ meV.
		Shades of green and blue denote the symmetry-broken metallic and full metal states, respectively.
		The symmetry-broken metallic states comprise spin-valley polarized (SVP), spin-polarized (SP), and spin-valley-locked (SVL) states.
		Specifically, SVP-I and SVP-II represent the $3/4$ and $1/4$ metal states, respectively.
		Inset schematics illustrate the Fermi contours for the four spin-valley flavors: $K\uparrow$, $K\downarrow$, $K^\prime\uparrow$, and $K^\prime\downarrow$.
	} 
\end{figure}
Next, starting from the SP state at $E=-6 \text{ mV/nm}$, an increase in hole concentration ($n<0$) activates the $K^\prime\uparrow$ channel. This new contribution to the conductivity drives a transition into a type-I SVP state. Although this phase is the same $3/4$-metal phase as SVP at the $n>0$ side (top-right of Fig. \ref{fig:10}), its total conductivity is primarily built upon the conducting channels of the parent SP state, rather than the SVL state.
Furthermore, a type-II SVP state, equivalent to a 1/4-metal phase with only one channel is conducting and the remaining three are insulating, was also identified.

Overall, for the symmetry-broken metallic states highlighted in the light green regions of Fig. \ref{fig:10}, the linearly dispersing bands do not participate in electrical conduction, suggesting a relative stability in the Fermi surface topology and charge transport characteristics, as the involvement of linear bands would typically yield minimal contributions to the total conductivity due to their intrinsically low DOS.

Given the prediction of these metallic states at fractional fillings, it is natural to ask whether they indicate a pathway towards achieving fractional quantum anomalous Hall (FQAH) phases.
This pathway, however, is obstructed by a fundamental obstacle: the absence of an energetically isolated topological band.
This condition is met in moir\'e superlattices~\cite{lu_fractional_2024} but fails in our non-moir\'e system, where inherent band overlap prevents the precise fractional filling required to open an FQAH gap.

\section{\label{sec:level1}CONCLUSIONS}
In summary, our theoretical investigation demonstrates the emergence of a QAH state in ABCB.
This arises from the intricate interplay of its intrinsic spontaneous polarization, strong correlations, and SOC.
Notably, we find that substantial $U=8 \text{ eV}$, in conjunction with the inherent polarization and SOC, are sufficient to drive a $C=3$ QAH state at $E=0$.
This topological phase also exhibits clear tunability: for instance, moderate $U=6 \text{ eV}$ necessitates a minimal electric field to stabilize the $C=3$ Chern insulator, suggesting a cooperative mechanism.
Beyond the QAH insulator, our calculations predict a rich landscape of other correlation- and SOC-induced symmetry-broken metallic states, including quarter- and three-quarter-filled phases, accessible by $E$ and $n$.

It is pertinent to acknowledge our model's reliance on phenomenological interaction parameters. 
Furthermore, it is important to recognize the limitations of the Hartree-Fock approximation used herein. As a mean-field theory, it does not capture quantum fluctuations, which in strongly correlated systems can be significant. Consequently, while our model provides strong evidence for the emergence of a zero-field QAH state, its ultimate stability and the precise conditions for its appearance should be confirmed experimentally.
Meanwhile, the QAH gap in ABCB is small (1 meV) at $U=6 \text{ eV}$ and $\lambda=2 \text{ meV}$, especially when compared to the gap in ABCA.The experimental observation of such a small gap requires an ultra-clean environment. 
Recent advances in suppressing charge inhomogeneity via proximal graphite gates provide an experimental platform with low-disorder conditions, creating a more favorable environment for experimentally testing the predictions of our work.\cite{domaretskiy_proximity_2025}.
Furthermore, incorporating the effects of non-local (long-range) Coulomb interactions will be a crucial next step for a more comprehensive understanding of the phase competition.
Despite these complexities and the need for further refinement, these findings collectively underscore the rich physics at play.
This work shows that the confluence of intrinsic polarization, strong correlations, and SOC renders ABCB a highly tunable system for exploring crystalline-based topological phenomena.

Finally, we compare our predicted zero-field QAH state to twisted multilayer graphene. In moiré platforms like magic-angle twisted bilayer graphene, the QAH effect originates from engineered, ultra-narrow flat bands. This structure allows the Chern number (e.g., $C=1, 2, 3$) to be controlled by electrostatic doping, though lifting the requisite degeneracies may require a small external magnetic field\cite{wu_chern_2021}.In contrast, our moiré-less ABCB system operates on a different principle. Here, the topological character is an intrinsic property of the crystal's wide-bandwidth electronic structure, resulting in a fixed Chern number determined by the global correlated ground state that is not readily tuned by doping. The degeneracy lifting is entirely spontaneous, driven by electron interactions and SOC, which opens a global topological gap without an external field.

\appendix

\section{Determination of non-interacting ground state}\label{APP_A}
Evaluating theoretical tools for polarization-correlation coupling in ABCB is crucial, as methods have distinct merits and drawbacks.
Density Functional Theory (DFT) identifies zero-field non-interacting ground states but lacks an intuitive picture for interaction-driven physics and is computationally demanding for complex scenarios.
While effective tight-binding (TB) models, constructed by downfolding first-principles calculations via the Wannier projection method, can accurately reproduce DFT band structures, incorporating correlation effects and a magnetic field within this framework remains computationally intensive.
$\mathbf{k}\cdot\mathbf{p}$ models, expanded around high-symmetry points (e.g., $K$ point), provide a clear, efficient effective Hamiltonian for low-energy physics and iterative methods like Hartree-Fock.
However, reflecting its phenomenological nature, the k·p model's ground state predictions can deviate from DFT, making it better suited for explaining observations than for ab initio prediction.

To address these limitations, we integrate a modified  $\mathbf{k}\cdot\mathbf{p}$ effective model with self-consistent Hartree-Fock.
For predictive accuracy and physical clarity, DFT first determines the ABCB ground state.
Then,  $\mathbf{k}\cdot\mathbf{p}$ parameters—including hopping terms and all atomic on-site energies—are fine-tuned to match  $\mathbf{k}\cdot\mathbf{p}$ band structures and wavefunctions with DFT results.
This method combines $\mathbf{k}\cdot\mathbf{p}$ efficiency and clarity with a reliable DFT-benchmarked foundation for subsequent calculations.

\begin{figure}[htb] 
	\setcounter{figure}{0}
	\renewcommand{\thefigure}{A\arabic{figure}}
	\centering
	\includegraphics[width=0.5 \textwidth, trim=0 30 10 20,clip]{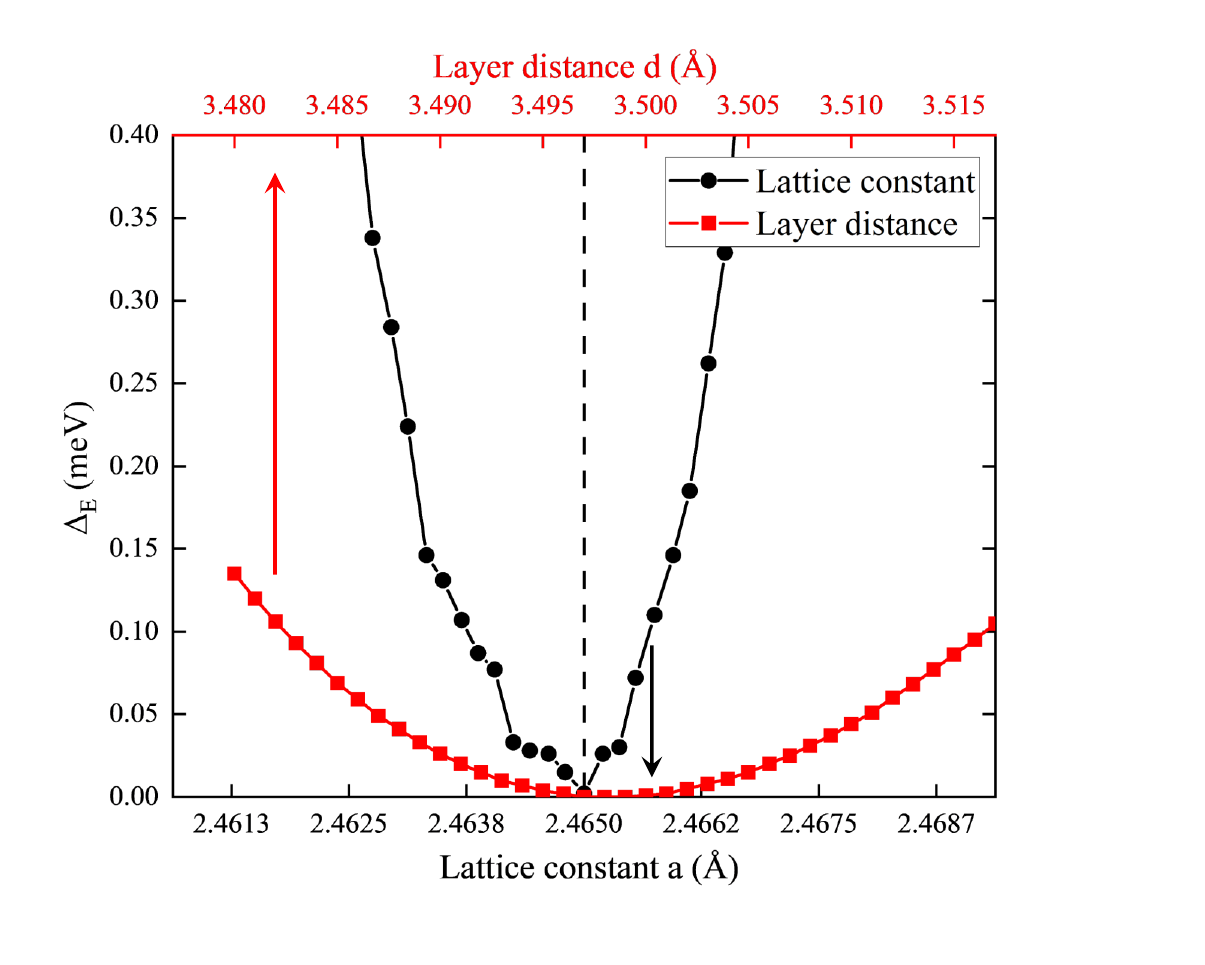}   
	\caption{\label{fig:a1}
		Result of DFT convergence test for ABCB, energy difference ($\Delta_E$, relative to the minimum total energy) versus lattice constant $a$ (black line with circles, bottom x-axis) and interlayer distance $d$ (red line with squares, top x-axis). The vertical dashed gray line indicates the optimized structural parameter where $\Delta_E$ is minimized.
	}    
\end{figure}
Our first-principles calculations were performed using the projector augmented-wave (PAW) method, as implemented in the Vienna Ab initio Simulation Package (VASP) \cite{kresse_efficient_1996,kresse_efficiency_1996,kresse_ab_1993}. For the exchange-correlation energy, we employed the generalized gradient approximation (GGA) with the Perdew-Burke-Ernzerhof (PBE) functional \cite{perdew_generalized_1996}. Van der Waals interactions were also included via the DFT-D3 correction scheme \cite{grimme_consistent_2010,grimme_effect_2011}. For all computations, a plane-wave energy cutoff was set to 500~eV, and a vacuum layer of 20~\AA\ was added in the out-of-plane direction to avoid interactions between periodic images. To allow for potential symmetry-breaking, all symmetry operations were disabled.
The computational workflow proceeded in several stages. First, a structural relaxation was conducted using a $300\times300\times1$ $k$-point mesh, with the total energy converged to a precision of $10^{-7}$~eV. This procedure yielded an energetically favorable structure with an in-plane lattice constant of $a = 2.465$~\AA\ and an interlayer distance of $d = 3.497$~\AA (Fig. \ref{fig:a1}).
Next, for the subsequent calculations on this relaxed structure, a much denser $510\times510\times1$ $k$-point mesh was utilized for the self-consistent field calculation, with a Gaussian smearing of $10^{-3}$~eV and converging the total energy to within $10^{-8}$~eV.

\begin{figure}[htb]
	\renewcommand{\thefigure}{A\arabic{figure}}
	\centering  
	\includegraphics[width=0.35\textwidth, trim=20 20 20 10,clip]{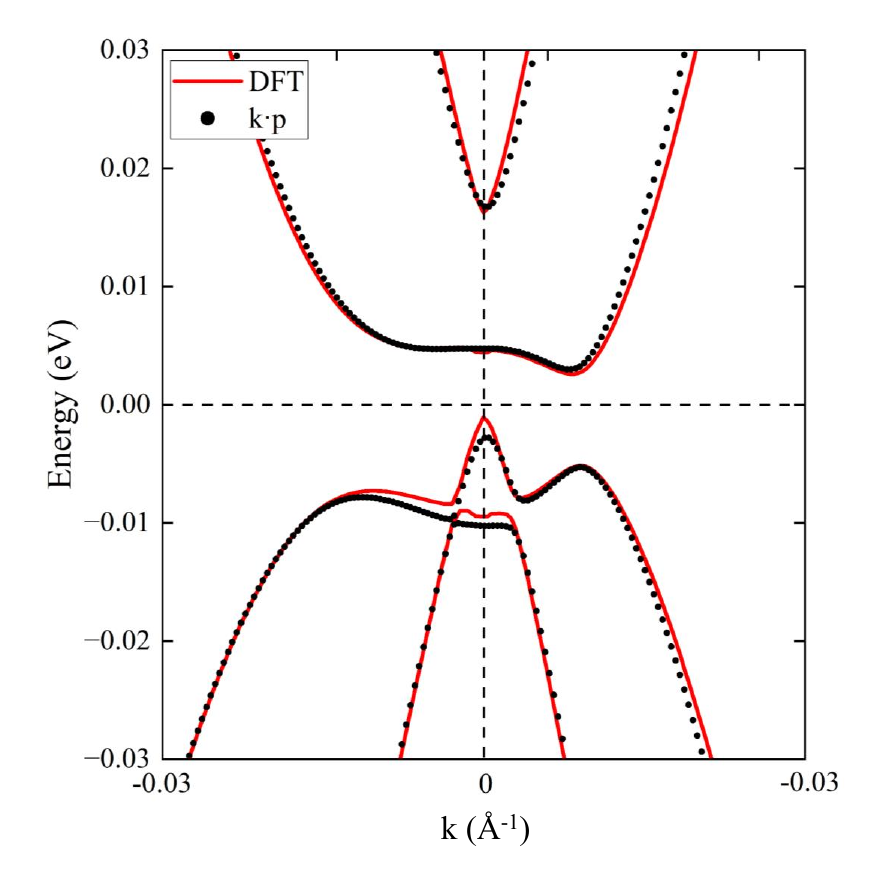}
	\caption{\label{fig:a2}
		Band structure comparison for ABCB, showing results from DFT (red lines) and the k·p model (black dots) around the high-symmetry-point K. The horizontal dashed line is the Fermi level.}
\end{figure}

\begin{figure*}[htb]
	\renewcommand{\thefigure}{A\arabic{figure}}
	\centering  
	\includegraphics[width=1\textwidth, trim=0 30 0 0,clip]{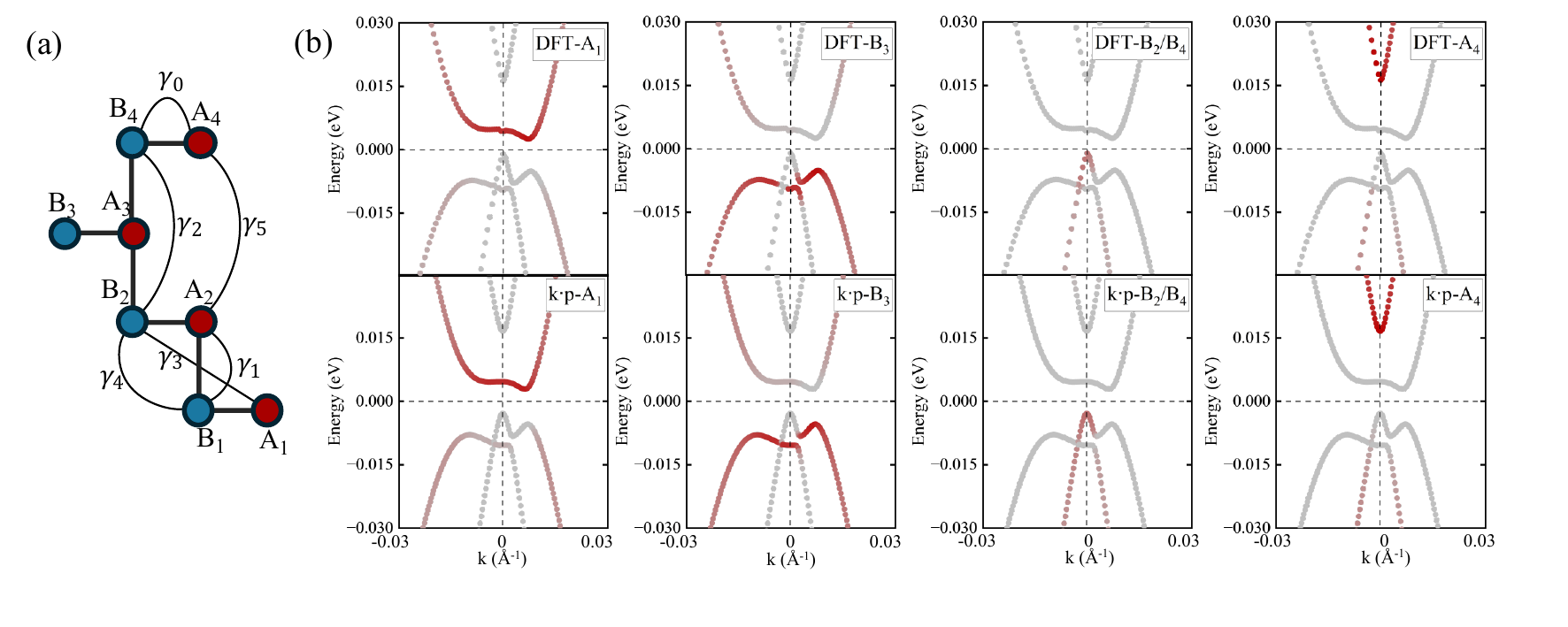} 
	\caption{\label{fig:a3}
		Schematic of the lattice model for ABCB and comparison of site-projected band structures from DFT and k·p calculations.
		(a) Illustration of the lattice sites (A$_1$/B$_1$ through A$_4$/B$_4$) and key intra- and inter-layer hopping parameters ($\gamma_0, \gamma_1, \ldots, \gamma_5$). 
		(b) Site-projected band structures from DFT (upper panels) and k·p model (lower panels).
		The sub-panels display projections onto sites A$_1$, B$_3$, B$_2$/B$_4$ and A$_4$ proceeding from left to right.
		}
\end{figure*}

By adjusting hopping parameters and on-site energies, the  $\mathbf{k}\cdot\mathbf{p}$ model yielded a band structure and wavefunction distribution consistent with DFT results.
This congruence is critical, as it circumvents scenarios where the  $\mathbf{k}\cdot\mathbf{p}$ band structure aligns with DFT, yet the wavefunction distribution is contrary, a key consideration for accurately determining the ground state using the  $\mathbf{k}\cdot\mathbf{p}$ model.
The resultant findings are illustrated in Fig. \ref{fig:a2} and \ref{fig:a3}.

\section {Self-Consistent Hartree-Fock Method}\label{APP_B}

The interacting Hamiltonian is solved self-consistently within the Hartree-Fock approximation. The effective Hamiltonian in momentum space takes the form:
\begin{equation}
	h_{\text{HF}}(\bm{k}) = h_0(\bm{k}) + \Sigma[\mathbf{P}]
\end{equation}
where $h_0(\bm{k})$ is the non-interacting part, and the self-energy $\Sigma$ is a functional of the density matrix $\mathbf{P}$. The density matrix is defined as $P_{\alpha, \beta}^{i}(\bm{k}) = \langle c_{i,\bm{k},\alpha}^{\dagger}c_{i,\bm{k},\beta}\rangle$, where the expectation value is taken with respect to the ground state of $h_{\text{HF}}(\bm{k})$.

The self-energy $\Sigma$ consists of the standard Hartree and Fock terms, which depend on the interaction parameters $U$, $V$, and $J_H$. A detailed derivation of these terms for a rhombohedral graphene can be found in the work\cite{liu_layer-dependent_2025}.

The self-consistent loop is iterated until the change in the total ground state energy between successive steps is less than $10^{-7}$~eV. Within the parameter range studied, we confirmed that the calculation converges to the same lowest-energy ground state regardless of the initial trial state. A linear mixing parameter of $\alpha=0.3$ is used to ensure numerical stability. The Brillouin zone is sampled on a $90 \times 90$ mesh around the $K$ and $K^{\prime}$ valleys with a momentum cutoff of $\Lambda = 0.25/a_0$, where $a_0$ is the graphene lattice constant\cite{jung_lattice_2011}. 

\section{Choice of the Hubbard Interaction Parameter U}\label{APP_C}
In our mean-field study of statically ordered ground states in ABCB, the on-site Hubbard interaction $U$ is a phenomenological effective parameter ($U_{\text{eff}}$). This parameter must account for screening effects from high-energy bands not included in our low-energy model, and is conceptually distinct from the smaller, partially screened $U$ used in many-body fluctuation theories.

To establish a physically meaningful range for $U_{\text{eff}}$, our choice is grounded in foundational \textit{ab initio} calculations that determine the partially screened Coulomb interactions in graphene. A pioneering study by Wehling et al.\cite{wehling_strength_2011} employed the constrained Random Phase Approximation (cRPA) to compute the effective interactions for a low-energy model. They determined the on-site Hubbard interaction for freestanding graphene to be $U = 9.3~\text{eV}$. Crucially, this value places graphene in a moderately correlated regime, close to the quantum phase transition towards a spin-liquid insulating state. This work was later corroborated and extended by Rösner et al.\cite{rosner_wannier_2015}, who performed independent cRPA calculations yielding a consistent value of $U = 9.7~\text{eV}$ for monolayer graphene. Furthermore, their newly developed and highly efficient Wannier function continuum electrostatics (WFCE) method also produced a value in agreement, $U \approx 9.8-10.0~\text{eV}$.

Collectively, these seminal works establish a well-justified \textit{ab initio} range of approximately $9.3-10.0~\text{eV}$ for the effective on-site Coulomb interaction in freestanding graphene. It is worth noting, however, that the presence of substrates and gate dielectrics in realistic device geometries provides additional screening that can reduce $U_{\text{eff}}$. Therefore, our investigation of the system's behavior for $U$ in the range of 6 to 9 eV represents a systematic exploration of the parameter space from the moderately correlated regime up to the strongly correlated limit, while also covering the physically relevant range for typical experimental setups.

\section{Interaction Parameters $V$ and $J_H$}\label{APP_D}
\begin{figure}[htb] 
	\renewcommand{\thefigure}{A\arabic{figure}}
	\centering
	\includegraphics[width=0.48 \textwidth, trim=70 20 60 20,clip]{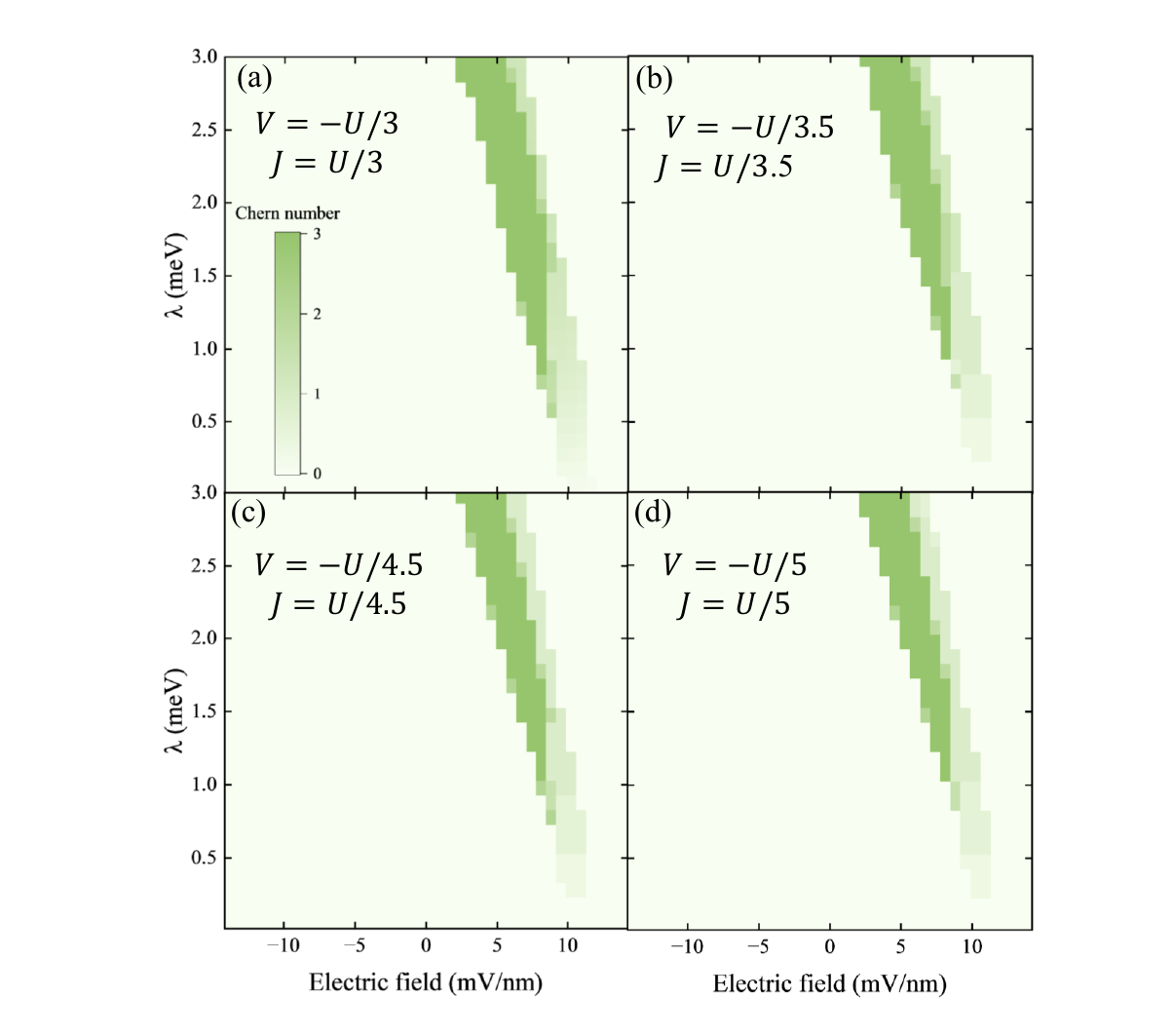}    
	\caption{\label{fig:a4} Phase diagram of ABCB of the Chern number as a function of SOC strength $\lambda$ and electric field $E$. All calculations are performed with a fixed on-site interaction $U = 6$ eV. (a)-(d) show the evolution of the QAH phase boundary for varying strengths of the non-local interactions: (a) $V = -U/3$, $J = U/3$; (b) $V = -U/3.5$, $J = U/3.5$; (c) $V = -U/4.5$, $J = U/4.5$; and (d) $V = -U/5$, $J = U/5$.}    
\end{figure}

In the main text, our analysis primarily employs the representative parameterization $V = -U/4$ and $J_H = U/4$. This choice is physically motivated by the common origin of the on-site interaction terms from the underlying screened Coulomb potential. To confirm that our main conclusions are robust and not an artifact of this specific ratio, we have systematically investigated the system's behavior across a range of non-local interaction strengths.
As shown in Fig.~\ref{fig:a4}, the QAH phase boundary shifts only slightly as the ratios of the non-local interactions, $-V/U$ and $J_H/U$, are varied from 1/3 to 1/5. This weak dependence on the specific parameterization confirms that the key physical picture is robust and remains unaltered.

\section{Method: Landau Levels}\label{APP_E}
When a perpendicular magnetic field $\mathbf{B} = (0,0,B_z)$ is applied, we adopt the Landau gauge, $\mathbf{A} = (-B_z y, 0, 0)$. The wave vector $\mathbf{k}$ is replaced by the canonical momentum operator $\hat{\mathbf{\pi}} = -i\hbar\nabla + e\mathbf{A}$.
The components of the canonical momentum operator, $\hat{\pi}_x$ and $\hat{\pi}_y$, satisfy the commutation relation $[\hat{\pi}_x, \hat{\pi}_y] = -i e B_z \hbar = -i \hbar^2/l_B^2$, where $l_B = \sqrt{\hbar/(eB_z)}$ is the magnetic length. We can define ladder operators $\hat{a}$ and $\hat{a}^\dagger$:
\begin{equation}
	\hat{a} = \frac{l_B}{\sqrt{2}\hbar}(\hat{\pi}_x - i\hat{\pi}_y),  
	\hat{a}^\dagger = \frac{l_B}{\sqrt{2}\hbar}(\hat{\pi}_x + i\hat{\pi}_y)
\end{equation}

We define the operators $\pi = \hat{\pi}_x - i\hat{\pi}_y$ and its Hermitian conjugate $\pi^\dagger = \hat{\pi}_x + i\hat{\pi}_y$.
\begin{equation}
	\pi = \frac{\sqrt{2}\hbar}{l_B} \hat{a}, \qquad \pi^\dagger = \frac{\sqrt{2}\hbar}{l_B} \hat{a}^\dagger
\end{equation}
The ladder operators $\hat{a}, \hat{a}^\dagger$ act on the harmonic oscillator states $|n\rangle$ ($n=0, 1, 2, \dots, N_{cut}$) as:
\begin{equation}
\begin{aligned}
	\hat{a}|n\rangle &= \sqrt{n}|n-1\rangle \\
	\hat{a}^\dagger|n\rangle &= \sqrt{n+1}|n+1\rangle \\
	\hat{a}|0\rangle &= 0
\end{aligned}
\end{equation}
This basis $\{|n\rangle\}$ describes the quantized cyclotron orbits of electrons in the magnetic field.
Each matrix element $H_{0,ij}$ of the original $N_{orb} \times N_{orb}$ $\mathbf{k}\cdot\mathbf{p}$ Hamiltonian $H_0$ is now expanded into an $(N_{cut}+1) \times (N_{cut}+1)$ block matrix $(H_{B})_{ij}$.

$H_{0,ij}$ is the matrix element of $H_{0}$ without magnetic field, if $H_{0,ij}$ is a constant $C_{ij}$ independent of $k$, the corresponding block matrix is $C_{ij} \cdot \mathbb{I}$, where $\mathbb{I}$ is the $(N_{cut}+1) \times (N_{cut}+1)$ identity matrix.
If $H_{0,ij}$ depends linearly on $\pi$ or $\pi^\dagger$ (e.g., $v_0 \pi^\dagger$), the corresponding block matrix is $v_0 \frac{\sqrt{2}\hbar}{l_B} \cdot M_a$ or $v_0 \frac{\sqrt{2}\hbar}{l_B} \cdot M_{a^\dagger}$, where $M_a$ and $M_{a^\dagger}$ are the matrix representations of the operators $\hat{a}$ and $\hat{a}^\dagger$ in the $\{|n\rangle\}$ basis, respectively.
\begin{equation}
\begin{aligned}
	(M_a)_{n',n} &= \langle n' | \hat{a} | n \rangle = \sqrt{n} \delta_{n',n-1} \\
	(M_{a^\dagger})_{n',n} &= \langle n' | \hat{a}^\dagger | n \rangle = \sqrt{n+1} \delta_{n',n+1}
\end{aligned}
\end{equation}
The final Hamiltonian $H_{B}$ is a $N_{orb}(N_{cut}+1) \times N_{orb}(N_{cut}+1)$ matrix.

\section{Topological Non-trivial States of ABCA}\label{APP_F}
\begin{figure}[htb] 
	\renewcommand{\thefigure}{A\arabic{figure}}
	\centering
	\includegraphics[width=0.45 \textwidth, trim=30 30 30 0,clip]{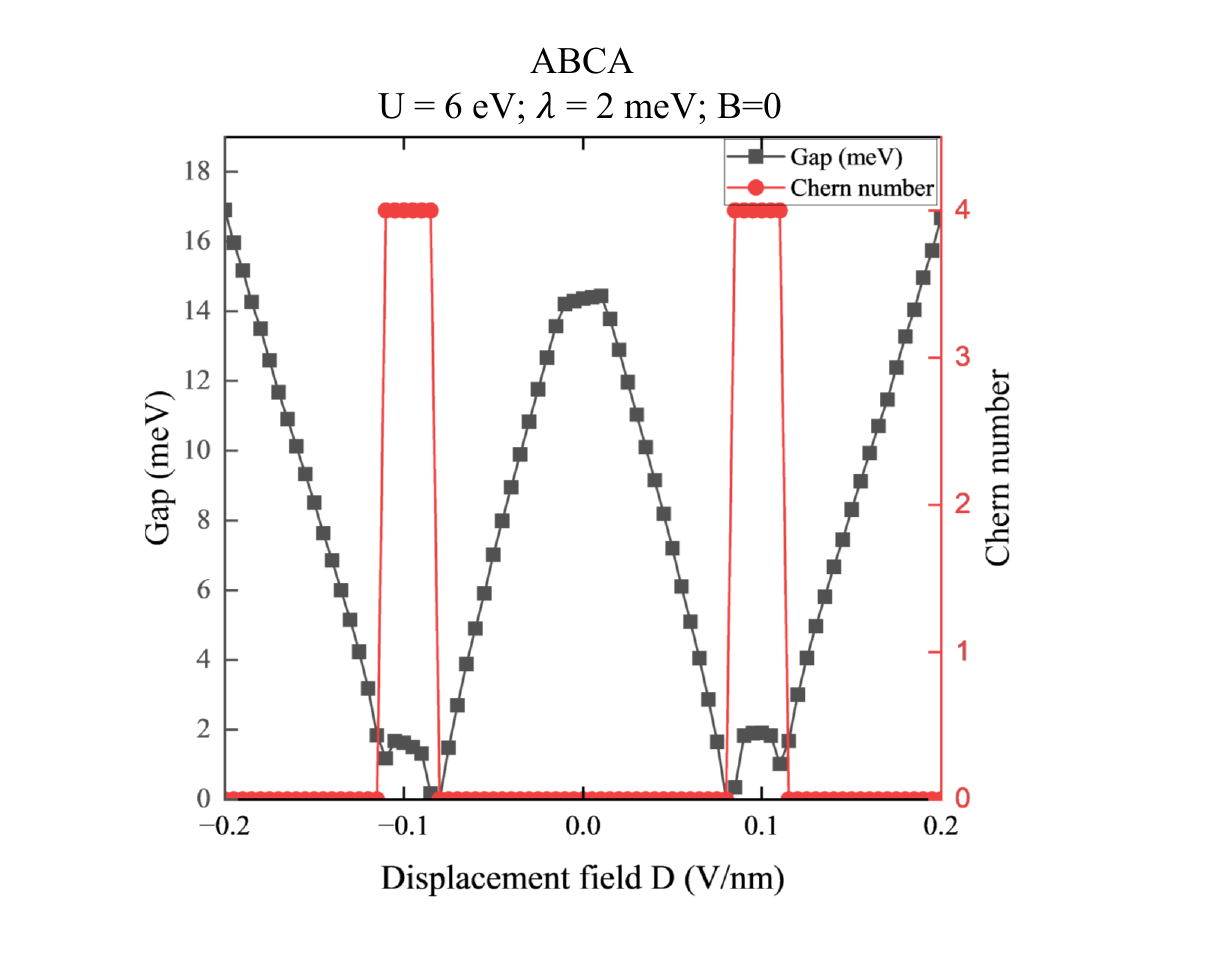} 
	\caption{\label{fig:a5}
		Calculated energy gap (black squares, left axis) and Chern number (red circles, right axis) as a function of displacement field $D$ for ABCA. The calculation was performed with an on-site interaction $U = 6$ eV and a spin-orbit coupling $\lambda = 2$ meV. For comparison with experimental results, an effective dielectric constant of $\epsilon_r=4$ is used.
	} 
\end{figure} 

\begin{figure}[htb] 
	\renewcommand{\thefigure}{A\arabic{figure}}
	\centering
	\includegraphics[width=0.35 \textwidth, trim=0 10 0 10,clip]{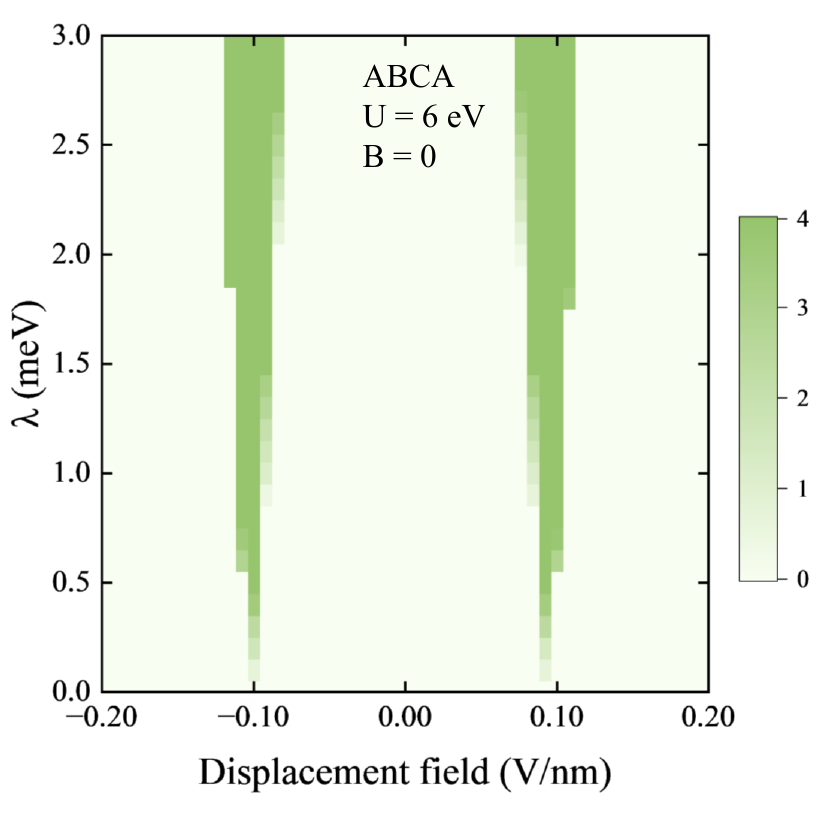}   
	\caption{\label{fig:a6}
		Chern number phase diagram for ABCA as a function of $\lambda$ and displacement field. The calculation is for a fixed interaction $U = 6$ eV, using an effective dielectric constant of $\epsilon_r = 4$.
	}
\end{figure}

To validate our theoretical model and choice of parameters, we first perform benchmark calculations on the well-studied ABCA. 
Figure~\ref{fig:a5} plots the calculated energy gap and Chern number as a function of the out-of-plane displacement field $D$ for an interaction strength of $U=6$~eV and a SOC of $\lambda=2$~meV.


Our calculations establish that at zero displacement field, electron-electron interactions drive the ground state of ABCA into a LAF insulator with an energy gap of approximately 14.5 meV, a finding that aligns with the work of Liu et al.~\cite{liu_spontaneous_2024}. Furthermore, as a displacement field $D$ is applied, the subsequent evolution of the gap—mirroring the experimentally measured longitudinal resistance ($R_{xx}$)—the topological transition at a critical field of $|D_c| \approx 0.1$ V/nm, and the emergence of a $C=4$ QAH state are all in quantitative agreement with the experiments by Sha et al.\cite{sha_observation_2024}. In addition, the reliability of this theoretical approach is further underscored by recent work on pentalayer rhombohedral graphene, where similar on-site interaction models have also reproduced experimental QAH states\cite{liu_layer-dependent_2025}.


This field-driven band inversion mechanism, where a single spin-valley flavor pair inverts to produce a QAH state, is also fully consistent with the schematic physical picture for ABCA presented in the main text's Fig.~\ref{fig:1}.

Figure~\ref{fig:a6} maps out the phase diagram as a function of the displacement field $D$. For a representative SOC strength of $\lambda = 2$~meV, the resulting QAH window is one-side $D \in [-0.11, -0.08]$~V/nm. With an effective dielectric constant of $\epsilon_r = 4$, this corresponds to the both-side field $|E|$ range of $E \in [20, 27.5]$~mV/nm for ABCA, as cited in the main text.

\section{Topological Non-trivial States of ABCA}\label{APP_G}

\begin{figure}[htb] 
	\renewcommand{\thefigure}{A\arabic{figure}}
	\centering
	\includegraphics[width=0.5 \textwidth, trim=10 0 0 0,clip]{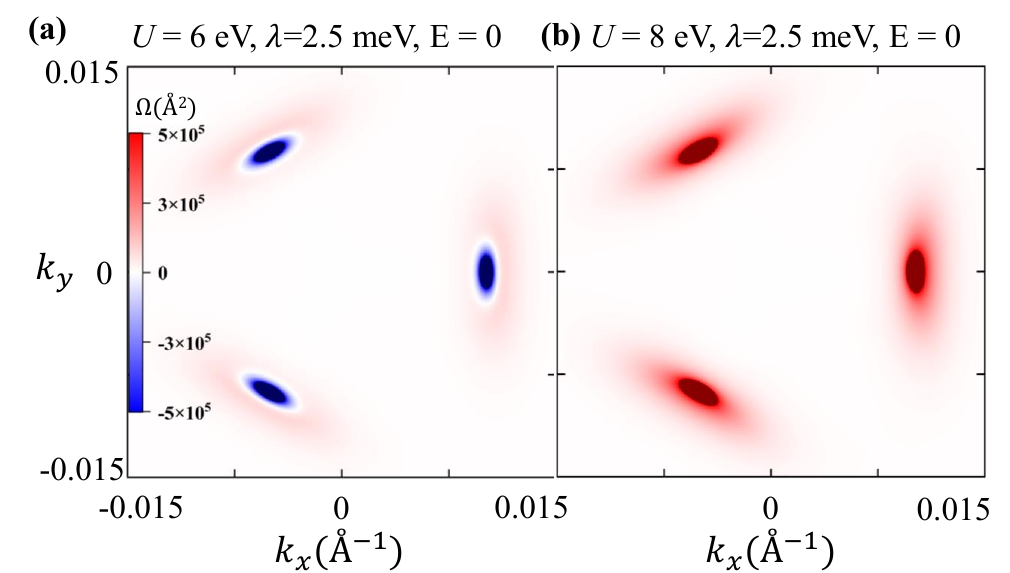}    
	\caption{\label{fig:a7}Total Berry curvature distribution in ABCB, obtained by summing over all four spin–valley channels, for (a) $U = 6~\mathrm{eV}$ and (b) $U = 8~\mathrm{eV}$, with $\lambda = 2.5~\mathrm{meV}$ and $E = 0$.}
\end{figure}

The Chern numbers $C = 3$ in the ABCB stacking and $C = 4$ in the ABCA stacking were obtained via Berry curvature analysis. The Chern number is calculated by integrating the Berry curvature $\Omega(\mathbf{k})$ of all occupied states over the Brillouin zone (BZ):  
\[
C = \frac{1}{2\pi} \int_{\mathrm{BZ}} \Omega(\mathbf{k})\, d^2k
\]
as shown in Fig.~\ref{fig:a7}.  
For $U = 6\,\mathrm{eV}$ and a SOC strength of $2.5\,\mathrm{meV}$, the interaction is insufficient to induce band inversion in any channel. In this case, the total Berry curvature in momentum space exhibits both positive and negative components, and integration yields an overall Chern number of $C = 0$.  
When $U$ is increased to $8\,\mathrm{eV}$ with the SOC still at $2.5\,\mathrm{meV}$, the $K\!\downarrow$ channel undergoes band inversion. The total Berry curvature becomes entirely positive, and integrating it over the BZ results in $C = 3$. The value $C = 4$ in the ABCA stacking is obtained through an identical Berry curvature analysis.

\section*{Acknowledgments}
The authors thank Kai Liu and Guorui Chen for helpful discussions. This work was supported by the National Key R\&D Program of China (grant No.\ 2022YFA1402400). Computational resources were supported by the Center for High Performance Computing at Shanghai Jiao Tong University.

\bibliography{main}
	 
\end{document}